\documentstyle[11pt]{article}
\newcommand{\blankline}{\vskip .3cm}
\newcommand{\f}{\begin{equation}}
\newcommand{\ff}{\end{equation}}

\begin{document}
\rightline{\Large CGPG-95/4-5, IASSNS-95/29}
\centerline{\LARGE  Linking topological quantum field theory }
\blankline
\centerline{\LARGE   and}
\blankline
\centerline{\LARGE nonperturbative quantum gravity}
\rm
\vskip.3cm
\centerline{Lee Smolin${}^*$ }
\blankline
 \centerline{\it  Center for Gravitational Physics and Geometry}
\centerline{\it Department of Physics, The Pennsylvania 
State University}
\centerline{\it University Park, PA, USA 16802}
\centerline{and}
\centerline{\it   School of Natural Sciences, Institute for 
Advanced Study}
\centerline{\it Princeton, New Jersey, 08540, USA}
 \blankline
\centerline{ABSTRACT}
\noindent
\blankline
Quantum gravity is studied nonperturbatively in the case in which
space has a boundary with finite area.  A natural set of boundary 
conditions is studied in the Euclidean signature theory,  in which
the pullback of the curvature to the boundary  is self-dual (with a 
cosmological constant).  A Hilbert space which describes all the 
information accessible by measuring the metric and connection
induced in the boundary is constructed and is found to be the direct
sum of the state spaces of all $SU(2)$ Chern-Simon theories defined 
by all choices of punctures and representations on the  spatial 
boundary $\cal S$.  The integer level $k$ of Chern-Simons theory 
is found to be given by $k= 6\pi /G^2 \Lambda + \alpha$, where 
$\Lambda$ is the cosmological constant and $\alpha$ is a $CP$ breaking 
phase.  Using these results, expectation values of observables which are 
functions of fields on the boundary may be evaluated in closed form.  The 
Beckenstein bound and 't Hooft-Susskind  holographic hypothesis  are confirmed, 
(in the limit of large area and small cosmological constant)  in the sense 
that once the two metric of the boundary has been measured,  the subspace of 
the physical state space that describes the further  information that the 
observer on the boundary may obtain about the interior  has finite dimension 
equal to the exponent of the area of the boundary, in Planck units, times a 
fixed constant.  Finally,the construction of the state space for quantum 
gravity in a region from that of all Chern-Simon theories defined on its 
boundary confirms the categorical-theoretic ``ladder of dimensions picture" 
of Crane.
\vfill
${}^*$ smolin@phys.psu.edu 
\eject
\section*{I. Introduction}

 In the last years significant progress has been made towards
the construction of a quantum theory of gravity in several
different directions.  Three of these directions, in particular,
have involved the use of new ideas and mathematical 
structures that seem,
in different ways, well suited to the problem of describing
the geometry of spacetime quantum mechanically.  These are
string theory\cite{strings}, topological 
quantum field 
theory\cite{witten-tqft,atiyah,segal,louis,louisdavid,louisigor}, and 
non-perturbative quantum gravity, based on the loop
representation
\cite{lp1,lp2,carlo-review,ls-review,aa-review,gangof5,jorgerodolfo}.   
Furthermore, despite genuine  
differences, there are a number of concepts
shared by these approaches, which suggests the possibility
of a deeper relation between them\cite{exp,discrete}.  
These include the 
common use 
of one dimensional rather than pointlike excitations, as
well as the appearance of structures associated with 
knot theory, spin networks and duality.  There are
also senses in which each development seems to lead to a
picture in which there is a discrete structure at short distances,
corresponding to there being only a finite number of degrees of
freedom per Planck volume of a system, or even per Planck area
of the boundary of the 
system
\cite{he,attickwitten,stringdiscrete,ks,ls-review,weave,volume}.

At the same time, each
development faces certain internal difficulties,  that have
so far resisted solution.  Futher,  none of these 
approaches has
been able to overcome
the great conceptual difficulties concerned
with extending the quantum description to the cosmological
case.   Because of this situation, it seems that it may be useful
to investigate the idea that a kind of unification of these different
approaches, taking what is successfully achieved by string
theory, topological quantum field theory and non-perturbative
quantum gravity, may be the right way to achieve a quantum
theory of gravity \cite{exp}.

In this paper I would like to propose one approach to
bringing together these separate developments.   
While still incomplete in certain aspects, this approach does show 
that by
incorporating  the methods of topological quantum field
theory into non-perturbative quantum gravity
certain things may be
achieved and the theory may be moved forward significantly.
Moreover, the key mathematical structure that makes this
possible turns out to be closely related to 
conformal field theory, which is the
basic basic mathematical framework for perturbative string theory.

The basic idea will be to study the quantization of the gravitational
field in a context in which we impose a certain kind of boundary
condition on spacetime.  This boundary will have a finite spatial
area, unlike the boundaries at infinity that are usually studied in
the asymptotically flat context in general relativity. 
The boundary condition will represent an idealized situation,
analogous to the case in which the electromagnetic field is
confined to a box.  We will assume that the observers, who live
outside of the box, can only observe the quantum gravitational
field in its interior by making measurements of fields it induced
on the walls.  

The main result will be that the quantum fields
that the observers who measure things at the walls have access
to may be described by topological quantum field theory.  Or,
more properly, by an infinite set of topological field theories.
For what we will find is that the state space of the gravitational
field of the interior is decomposed into the direct sum of an
infinite number of subspaces, each of which may be labeled
by eigenvalues of the operators that measure the surface areas
of  regions of the boundary.  
In each of these subspaces, further observations are described
in terms of a certain topological field theory.  One way to
say this is that in this formulation of quantum gravity, 
measurements of the metric geometry of the boundary do not pick
out a state in a Hilbert space; they pick out a field theory, whose
states describe the possible further knowledge that the observer
may gain about the interior.

While we will derive these results completely from the methods
of canonical quantum gravity, they confirm the expectations of
two lines of thought, coming from topological quantum field theory
and string theory.

The first is the program Crane, who proposed a category-
theoretic formulation of quantum gravity motivated by the
problems of interpretation in quantum cosmology\cite{louis}.  
Basic to this 
proposal is a framework that Crane calls the ``ladder
of dimensions" that hypothesizes the existence of
certain  relationships tieing
together diffeomorphism invariant quantum field theories
in two, three and four 
dimensions\cite{louis,louisigor}.  
>From the mathematical
side, this framework has been so far realized in the constructions
of Crane and Yetter\cite{louisdavid} and 
Crane and Frenkel\cite{louisigor}.
We will see that non-perturbative quantum gravity with
the particular boundary conditions I mention here provides another
realization of this mathematical framework.  But as this
arises directly from the quantization of general relativity, this
may be said to confirm Crane's conjecture\cite{louis} that a theory of 
the
gravitational field involving an infinite number of degrees of 
freedom is in fact the right object to appear on the fourth rung of 
the ladder of dimensions.

One aspect of Crane's ladder of dimensions is that quantum gravity
in $3+1$ dimensions should be described, not by a single vector
space, but by a linear structure, which is spanned by basis elements
that 
each correspond themselves to a vector space.  
The idea is then that the vector spaces that may appear as the
``components of vectors" in the description of the 
$3+1$ dimensional theory are
the state spaces of appropriate $2+1$ dimensional theories.   
This is the basic
reason that category theory is necessary for this 
description, because it allows us to talk about a structure in which
one may take superpositions of vector spaces, or more simply to
describe vectors whose elements are 
themselves vector spaces\cite{other-categorists}.

This picture will be realized here in that the metric
of a spatial surface will turn out to label the different
topological quantum field theories that may be defined on it.   
The physical state space that describes the $4$ dimensional quantum 
gravitational field in a region bounded by that surface 
will then be constructed
from the state spaces of all the topological quantum field theories
that live on it.

This structure also provides a physically well defined framework
for the tangle algebra introduced by 
Baez in \cite{john-tangle}, and ties its
application in quantum gravity 
to its realization in topological quantum field theory.

The second development that these results support is the
holographic hypothesis that has arisin in the work 
of 't Hooft\cite{gerard-holographic}
and Susskind\cite{lenny-lorentz}.  
The basic idea of this hypothesis is that
a diffeomorphism invariant quantum field theory describing the
quantum geometry on the interior of a surface is best 
described by a quantum field theory on the boundary, 
rather than as a theory of local degrees of freedom in the
interior.  

Consider a region 
$\Sigma$ of space which is surrounded
by a spacial surface, ${\cal S} = \partial {\Sigma}$ with a finite 
area ${\cal A}[{\cal S}]$.  'tHooft and Susskind 
conjecture that in a quantum
theory of gravity the state space describing the physics in
${\Sigma}$ should be finite dimensional, 
with a dimension given by\cite{gerard-holographic,lenny-lorentz}
\f
d ({\cal S},h_{\alpha \beta})= e^{{\cal A}[{\cal S}]/cl_{Planck}^2},
\ff
where $h_{\alpha \beta }$ is the metric on the two surface
and $c$ is some
constant of proportionality of order one.  This is motivated
from two directions, first from the conjectured Beckenstein
bound\cite{beckenstein} on the information that 
can be contained within any
surface of finite area and second, from the behavoir of
string theory near horizons\cite{lenny-lorentz,lenny-bh}.

I may note that this motivation is greatly strengthened by the
recent derivation of Jacobson\cite{ted-new}, 
who shows, in 
essense, that the Beckenstein
bound must hold on the event horizons of all uniformally
accelerating observers if the Einstein equations are to hold
in the classical limit.

Here I will show that the Beckenstein bound and the
holographic hypothesis are, at least
under certain conditions and assumptions, consequences 
of non-perturbative
quantum gravity.   They may be derived under the
assumption that the space of states of the quantum
gravitaional field in the region ${\Sigma}$ must be spanned
by eigenstates of observables that are functions 
of fields on the boundary
${\cal S}$.  This would be the case, for example, if the surface
is just above a black hole horizon.  But, given that the boundary
has a finite area, this is plausibly the case generally.
Because of what we already have said, once the metric on the 
boundary is fixed, the possible states of the system that describe
further measurements of the gravitational field in the interior
that may be made by observers on the surface
are described in terms of a particular topological
field theory.  We will then find that the dimension of 
the state space of that theory
is, in the limit of small cosmological constant,
given by (1).

Before begining to describe these results in more detail,
I must mention their main limitations.  First, they have so far
been derived under a certain set of boundary conditions.
Whether there are other boundary conditions for which
similar results hold is as yet unknown.
Second, the results hold so far only in the case that the
relevant gauge group is compact.  This restricts us to the
case of Euclidean signature if we use the Ashtekar formalism.
 This is not necessarily
unphysical, as it may very well be that Euclidean quantum
gravity is the right context to discuss the theory at finite
temperature\cite{cstime}.
There is also another alternative which is to use instead the
new formalism of Barbero\cite{fernando-real}, in which
case general relativity with real Minkowskian signature
is described in terms of a real $SU(2)$ connection.
only in the case of Euclidean signature.  However, it is
also not impossible that it will be possible to continue
at least some of these results to 
Minkowskian signature in the Ashtekar formalism.
In the concluding section I will
discuss the prospects for overcoming both of these limitations.

As in many cases in nonperturbative quantum gravity, it is
easier to explain the main results then it is to understand
all the technical details involved in their derivation.  
Thus, I will start in the next
section 
with a sketch of the main features of topological quantum
field theory I will need, and then, in section $III$,
give a summary of the main
idea and results.  The derivation of these results are
given in the next three sections.  
Section $IV$ describes the classical theory with
the self-dual boundary conditions I will impose.  The classical
hamiltonian analysis of the theory with these boundary conditions
is the subject of section $V$, and the quantization, leading to the
main results, is given in section $VI$.  Some implications and
directions for further extension of these results are discussed
in the concluding section $VII$.

\section*{II. Summary of results from topological quantum field 
theory}

The basic picture of what a $3$ dimensional topological field theory
is was developed by Witten\cite{witten-tqft}, 
Atiyah\cite{atiyah} and 
Segal\cite{segal}.  I give here
a summary of the main ideas and results I will need.  I will
not give the most general or complete form of the theory, but
specialize to the case that will be of interest here, which is
the topological quantum field theory associated with $SU(2)$
Chern-Simons theory. 

The basic object that we will be concerned with is a compact
two dimensional surface $\cal S$, with a finite set of marked
points, $y_{\alpha }$, $\alpha = 1,...,n$.  These points
are also called punctures.  
Each point $y_\alpha$ is labled by   
half-integers $j_\alpha$ taken from the set $1/2 \leq j_\alpha \leq 
k/2$.
Here $k$, which is required to be an integer, is the coupling
constant, or level, of the Chern-Simons theory.

These $j_\alpha$'s label representations of
the quantum group $SL(2)_q$, with $q=e^{2 \pi \imath/k+2}$.
We may note that the
representations of quantum groups play an essential role in
topological quantum field theory; one result of the present
work will be the discovery of a role for them also in canonical
quantum gravity.  The fact
that there is a highest spin representation is a crucial property
of quantum groups (with $q$ at a root of unity) and this will play
a key role in the physical results such as the confirmation of the
Beckenstein bound.

The basic idea of a topological quantum field theory is that 
a finite dimensional 
vector space ${\cal H}_{{\cal S}, y_\alpha , j_\alpha}$
may be associated to  each set $({\cal S}, y_\alpha , j_\alpha )$.  
As these vector spaces will be the central objects on which
our theory is built, it will be useful to review some of their
properties.

Each state space may be considered to be the quantization
of $SU(2)$ Chern-Simons theory appropriate to the
spacial manifold ${\cal S}-y_\alpha $ with
sources at the marked points $y_\alpha$.  Chern-Simons
theory is described by the action,
\f
S_{CS}= {k \over 4 \pi } 
\int_{{\cal M}} Y_{CS}(a)
\ff
where $\cal M$ is a three dimensional manifold with
boundary ${\cal S}=\partial {\cal M}$
and $Y_{CS} (A) = {1\over 2}(A^i \wedge dA^i + {1 \over 3}
\epsilon_{ijk}
A^i \wedge A^j \wedge A^k) $ is the
Chern-Simons form\footnote{We use the notation
in which $a,b,c,...$ are spatial indices and $i,j,k$ are internal $SO(3)$
indices that label the frame fields of space.  
We will use units here in which $\hbar =1$, but $G$ is
written explicitly, so that $G$ has dimensions of $(length)^2$,
while the cosmological constant $\Lambda$ has dimensions
of $(length)^{-4}$. The combination $\lambda=G^2\Lambda $, where
$G$ is Newton's constant, is then dimensionless.}.   At the marked
points there are sources, so that the constraint
equation arising from (2) is
\f
F_{ }^i (\sigma ) =\sum_\alpha \lambda^i_\alpha 
\delta^2  (\sigma , y_\alpha )
\ff
where $\lambda_\alpha^i $ is a source, that lives in the
$j_\alpha$'th representation.  The treatement of the sources is
crucial to the correct quantization of Chern-Simons theory,
and will be discussed in Section VI.

To each such choice of surface, marked points
and representation is associated the vector space
${\cal H}_{{\cal S}, y_\alpha, j_\alpha }^{CS}$, which may be 
obtained from  the quantization of a phase space,
which is given by 
\f
{\cal P}^{CS} = { 
\mbox{Flat, SU(2) connections on } 
{\cal S} - \{ y_\alpha \}     \over
 SU(2)  \mbox{ gauge transformations }  }
\ff
with the restrictions (3).

I will denote by
$a_\alpha^i$ an $SU(2)$ connection on the two dimensional
manifold ${\cal S}-\{ y_\alpha \}$.   The basic
Poisson brackets are given by,
\f
\{ a_{\alpha}^i (\sigma ) , a_\beta^j (\sigma^\prime ) \}
= {2 \pi  \over k } \epsilon_{\alpha \beta } \delta^{ij} 
\delta^2 ( \sigma , \sigma^\prime )
\ff
We may note that the phase space ${\cal P}^{CS}$ is compact,
so that there are a finite number of independent observables.
A complete set of coordinates on the phase space is given
by the loop observables, 
\f
t[h(\alpha )] = TrP e^{\int_\alpha a}
\ff
which, by the constraints only depend on the homotopy class
of $\alpha$, which I denote by $h(\alpha )$.  These satisfy
an algebra, which is given by
\f
\{ t[h(\alpha )],  t[h(\beta )]    \} = {2 \pi  \over k} 
Int[h(\alpha)  , h(\beta ) ] \left (
t[h(\alpha ) \cdot h(\beta ) ] -t[h(\alpha ) \cdot h(\beta^{-1} ) ]
\right )
\ff
We will call the algebra of the observables of each of these 
Chern-Simons theory by
${\cal A}^{CS }_{{\cal S},y_\alpha ,j_\alpha}$.  A complete set of
these observables is given by the $t[\alpha ]$ for
$\alpha \in {\cal S}-\{ y_\alpha \}$.  
These also satisfy the reality conditions
\f
t[\alpha ]^* = t[\alpha ]
\ff
These observables then 
have representations on the finite dimensional
vector spaces ${\cal H}_{{\cal S},y_\alpha,j_\alpha}$
which preserves the reality conditions.   

In addition, we will need to take into account the
fact that the loop operator
$t[\alpha ]$ of a loop $\alpha $ that surround only one
point $y_\alpha $ are restricted by the condition (3).
We will see later in section VI
how this is accomplished. This will in fact
be the key point in the reduction of the dimensionality of the
state space to a finite value.  Roughly speaking, the space
of states ${\cal H}_{{\cal S}, y_\alpha, j_\alpha }^{CS}$ is 
spanned by a basis
that corresponds to the independent ways that the
spins $j_\alpha$ may be combined consistently according
to the rules of additional of angular momentum of the
quantum group $SL(2)_q$.   

Acting also on these states and observables are also
the  generators of the large diffeomorphism group
\f
Diff_{L}({\cal S}-y_\alpha ) =
{ Diff( {\cal S}-\{ y_\alpha \} ) \over 
Diff_{0}( {\cal S}-\{ y_\alpha \}) }
\ff
where $Diff_0$ denotes the component connected to the identity.

These representations are well understood
using the technology of conformal field 
theory\cite{witten-tqft,moreseiberg,cs-other,louis2d3d}. However,
as we will not
need the details of their construction here, I will
not describe them further.  I will need only
one fact, which is that in the limit of large $k$ the
${\cal H}_{{\cal S},y_\alpha ,j_\alpha}$  are 
given by \cite{witten-tqft}
\f
{\cal H}_{{\cal S},y_\alpha ,j_\alpha}
\rightarrow  Inv[\circ_\alpha R_{j_\alpha} ]
\ff
where $R_{j}$ is the spin $j$ representation of $SL(2)_q$ and
$Inv$ means the group invariant part.  As the taking of the group
invariant part involves a finite number of relations, we have,
in the limit of large $k$ 
\f
dim{\cal H}_{{\cal S},y_\alpha ,j_\alpha}  \rightarrow 
\prod_\alpha
(2j_\alpha +1 )
\ff
as, in the same limit, the dimensions of the quantum group 
representations
go over into their classical counterparts.

The second part of the definition of a topological
quantum field theory is a description of the states
in ${\cal H}_{{\cal S},y_\alpha,j_\alpha}$.  To describe
this we need to define a quantum spin network\cite{qnet}, 
which is a generalization
of the spin networks introduced by Penrose\cite{roger-spinnet}.

 A quantum spin network is
an oriented graph, which we denote also by $\Gamma$,
which is composed of smooth curve segments 
that meet at vertices.  Each line is labeled by an integer
$l $ between $0$ and $k/2$ denoting a representation
of $SL(2)_q$, and the rules for addition
of $q$-angular momentum must be satisfied at the vertices.
If there is more than one way for a singlet to be constructed
by the direct product of the representations given by the
labels of the lines entering a vertex, then that vertex
must be labeled by an index labeling these different possibilities.
There is no additional labeling for trivalent vertices.

According to the Atiyah axioms, it is the case that for
every choice of compact
manifolds
$\Sigma$ whose boundary is $\cal S$, together with a quantum
spin network $\Gamma$ in $\Sigma$, that meets the boundary
at the marked points $y_\alpha$, so that the labels on the lines
that go into the point match their labels $j_\alpha$,   
there is a quantum state in ${\cal H}^{CS}_{{\cal S}, 
y_\alpha,j_\alpha}$.
I will denote this state 
$|\Gamma , \Sigma >^{CS} \in {\cal H}^{CS}_{{\cal 
S},y_\alpha,j_\alpha}$. 
(Note that in cases where $\Sigma$ is fixed it will not be
explictly indicated, so these states are simply $|\Gamma >^{CS}$.) 
We may note that this association of manifolds with imbedded
$q$-spin networks and states is highly non-unique, as the
dimensionality of each of the state spaces is finite dimensional,
while there are an infinite numbers of manifolds and graphs that
can match the boundary.  It is essentially because of this
non-uniqueness that this construction yields a useful recursive
topological invariant.

There is also a duality relation on the 
${\cal H}_{{\cal S},y_\alpha,j_\alpha}$ that reverses
the orientation of $\cal S$ and takes each representation
$j_\alpha$ into its dual.  Given this there is an inner product
on states in which $<\Gamma ,\Sigma |\Gamma^\prime 
,\Sigma^\prime >^{CS}$ 
is given by
a formal path integral by the following construction. 
A closed quantum spin network in the manifold
$\Sigma \# \Sigma^\prime$, gotten
by joining then along $\cal S$,  
can be constructed by
joining the ends of $\Gamma$ and $\Gamma^\prime$ at the marked
points of $\cal S$, so that the representations on the joined edges
coincide.  
We may then define
\f
<\Gamma ,\Sigma |\Gamma^\prime ,\Sigma^\prime >^{CS}=
\int d\mu (a) 
e^{{k \over 4 \pi } 
\int_{{\cal M}} Y_{CS}(a)} T[\Gamma \circ \Gamma^\prime ]
\ff
where $T[\Gamma \circ \Gamma^\prime ]$ is the appropriate 
product
of traces of holonomy of the $SU(2)$ connection $a$ and the integral
is over connections in the manifold $\Sigma \# \Sigma^\prime $.

To define the integral the holonomies of the loops must be
framed\cite{witten-tqft}.  Once this is done, this 
may be evaluated using either
conformal field theory as described by Witten \cite{witten-tqft} 
or by explictly
evaluating the integral, using perturbation 
theory\cite{ls-cs,cs-pert}.  The result is,
\f
<\Gamma ,\Sigma |\Gamma^\prime ,\Sigma^\prime >={\cal K}^k 
[\Gamma \circ \Gamma^\prime , \Sigma \# \Sigma^\prime ]
\ff
which is the Kauffman bracket of the spin network 
$\Gamma \circ \Gamma^\prime$
in the 
compact manifold $\Sigma \# \Sigma^\prime $.

Finally, this construction can also be used to construct
linear maps between two different state spaces,
${\cal H}_{{\cal S},y_\alpha,j_\alpha}$, associated with two
marked punctured surfaces $({\cal S},y_\alpha , j_\alpha )$
and $({\cal S^\prime },y_\alpha^\prime  , j_\alpha^\prime )$.  For
every way of choosing a compact manifold $\Sigma$, whose 
boundary
is given by ${\cal S} \cup {\cal S}^\prime $, and for every
q-spin network $\Gamma$ inbedded in it that meets each boundary
at the marked points, matching the representations, there is
a corresponding linear map,
\f
{\cal M}_{\Gamma , \Sigma } : 
{\cal H}_{{\cal S},y_\alpha,j_\alpha}^k
\rightarrow 
{\cal H}_{{\cal S}^\prime,y_\alpha^\prime ,j_\alpha^\prime}^k .
\ff

\section*{III. The basic idea}

Now I will describe the basic idea of using the 
structure of Chern-Simons
theory to describe a set of observables in
non-perturbative quantum gravity. In this section 
I sketch and motivate
the construction, in later sections the key mathematical relations
that are needed to realize this idea are derived from quantum 
general
relativity by imposing an appropriate boundary condition.

Let us assume that we are a family of observers, for whom there is
a region of spacetime, $\cal M$ that we cannot directly observe. 
This region has a boundary $\partial {\cal M}$ that marks the limits
of where we may probe with our measuring instruments.  We will
assume that ${\cal M} = \Sigma \times R$, where $R$ will be the 
time
direction, or that ${\cal M}=\Sigma \times S^1$, in the case we
want to do finite temperature quantum physics.  It then follows
that the boundary is $\partial {\cal M}=\partial \Sigma \times R$
or $\partial {\cal M}=\partial \Sigma \times S^1$, respectively.
As in the last section, $\partial \Sigma$ will be denoted by
$\cal S$, and will be assumed to be compact without boundary.

There are two kinds of situations in which this may be imagined
to occur.  It could be that we have constructed walls out of a very
special material that is perfectly reflecting to gravitational waves,
in which case $\cal S$ denotes the position of the walls.  We are then
able to position detectors just at the walls, but no further inside.
(We may note that because of the results 
of \cite{intrinsic-entropy} the walls must
be constructed with material that violates the positive energy
conditions, but perhaps we can do this by cleverly manipulating the
vacuum of a gauge theory.)

The second case that could be relevant is that in which
$\partial {\cal M}$ is a black hole or cosmological event horizon 
or a horizon that is present because we are a 
family of uniformly accelerating 
observers.
In this case the spacial boundary $\cal S$ corresponds to the
intersection of the null horizon with some three dimensional surface
of simultaneity.

In either case we have a system, the gravitational field in the
interior of $\Sigma$, whose quantization we will now consider, using
the methods of nonperturbative quantum gravity.

To proceed we must say something about the gauge invariances and 
symmetries
on the boundary $\cal S$.  The simplest assumption, which we will 
see
below can be realized explicitly, is that in which 
we can realize the $SU(2)$
symmetry on the full $\Sigma$, including the boundary, but all
diffeomorphisms are constrained to vanish on the 
spacetime boundary $\partial {\cal M}$.  In this case, neither the
diffeomorphism or hamiltonian constraints can generate gauge 
transformations
at the boundary. 

Once we have fixed the gauge transformations on the boundary,
we can ask what observables are accessible to our
family of observers who cannot penetrate inside the boundary
$\partial {\cal M}$.  Given that there are no diffeomorphisms
on the boundary, we may note that the following two sets of
observables will be well defined on the boundary:

${\cal A}[{\cal R}]$, the area of any finite region $\cal R$ in the
spatial boundary $\cal S$.

$T[\alpha ]$, the Wilson loop of the
Ashtekar connection of any loop $\alpha \in {\cal S}$.

We may note that by the previous 
results\cite{volume} each area commutes
with all the other observables in this set.  However, we will see
that we cannot assume that the different Wilson loops on the 
boundary commute with each other.  We may note also that as
the diffeomorphisms are broken at the surface, different
regions and loops
on $\cal S$ are distinct.  The diffeomorphisms of the spatial
boundary, denoted,  
$Diff({\cal S})$, are not gauge transformations,
instead they comprise a symmetry group.  
Its generators, $D(v)$, where $v^\alpha$ will denote vector
fields on $\cal S$, must then also be observables.

We will denote the algebra of observables generated by these
three sets, ${\cal A}[{\cal R}]$,
$T[\alpha ]$ and $D(v)$ by ${\cal A}^{boundary}_{\cal S}$.  These
are, by assumption, what we as observers unable to penetrate
into the interior of $\Sigma$ can measure.  They comprise
a subalgebra of the full algebra of physical observables 
of the gravitational field in the
manifold $\cal M$, which I will denote by ${\cal 
A}^{phys}_{\Sigma}$.

Let us then see how far we can get by applying the results of
non-perturbative quantum 
gravity\cite{lp1,carlo-review,ls-review,weave,volume} 
to this situation.  We
will try to use these observables to learn as much as possible
about the physical state space ${\cal H}^{phys}_{\Sigma}$ 
that describes
the quantum gravitational field in the interior of the region
$\Sigma$.  We will begin, however, by studying the simpler problem
of characterizing the larger space of spatially diffeomorphism
invariant states, which we will denote, ${\cal H}^{diffeo}_{\Sigma}$

As they commute among themselves and 
with the Wilson loops, and transform naturally
among themselves under the $D(v)$, it is natural to begin to 
characterize
the space in terms of the eigenstates of the area operators
${\cal A}[{\cal R}]$.  We may then seek to use immediately the
results\cite{volume} that the 
eigenstates of the operators that measure the
area of a region in quantum gravity are the spin network
states\cite{spinnet-us}.

As the diffeomorphisms are broken at the boundary, but gauge 
invariance
is preserved, we may expect that the diffeomorphism and gauge 
invariant
states are described by spin networks $\Gamma$ in $\Sigma$, which 
are
distinct up to diffeomorphisms that leave the boundary fixed.  I
will denote such states by $|\Gamma>^{gr}$.  
We may note that the
spin networks $\Gamma$ here may meet the boundary and then run 
along
inside of it.  Each spin network $\Gamma$ then has associated to it
a set of marked points, $y^\Gamma_\alpha$, where it meets the 
boundary
$\cal S$ and labels $j_\alpha^\Gamma$ of the spins of the lines that 
meet
the boundary.

Note that as no line can end, in the boundary or elsewhere, the total
spin entering the boundary must vanish.

At this stage, we are dealing with the ordinary spin networks labeled
by representations of $SU(2)$, in contrast to the quantum spin
networks that play a role in Chern-Simons theory.

As we noted, the states $|\Gamma >^{gr}$ are all eigenstates
of the area operators $\hat{\cal A}[{\cal R}]$.  The corresponding
eigenvalues defined by 
\f
\hat{\cal A}[{\cal R}] |\Gamma >^{gr} = a({\cal R},\gamma ) 
|\Gamma >  ,
\ff
are given by\cite{volume}
\f
a({\cal R},\gamma )=\sum_{y^\Gamma_\alpha \in {\cal R}}
{l_{pl}^2 \over 2}
\sqrt{j_\alpha (j_\alpha +1 )}  .
\ff
The factor of $1/2$ in this expression comes from the fact that
the edges of $\Gamma$ do not cross the boundary $\cal S$, but meet
it and then run along in it. The intersection number of each with
the boundary is then $1/2$.

We may then decompose the state space of quantum gravity 
according
to degenerate eigenspaces of these area observables.
Thus, for each set of marked points $y_\alpha$
and representation labels $j_\alpha$ on $\cal S$ there 
is a subspace
${\cal H}^{diffeo}_{y_\alpha ,j_\alpha}$ of 
${\cal H}^{diffeo}_{\Sigma}$,
spanned by the states $|\Gamma >^{gr}$
where the $\Gamma$ are the spin networks that meet 
the boundary at those points,
with lines colored with those representations.  Each state
$|\Gamma >^{gr}$ must be in one of those suspaces, so that
\f
{\cal H}^{diffeo}_{\Sigma}= \sum_{n=0}^\infty 
\sum_{j_1...j_n }\int d^2y_1...d^2y_n \ 
{\cal H}^{diffeo}_{y_\alpha ,j_\alpha}
\ff

Furthermore, we know that in any physical inner product,
states in different of these subspaces must be orthogonal
to each other, as they have different eigenvalues of a
physical observable.

We may note that some of the observables in ${\cal 
A}^{boundary}_{\cal S}$
mix up these subspaces, while others leave them invariant.  
To describe this situation more precisely, let
${\cal A}^{gr}_{{\cal S}, y_\alpha , j_\alpha}$ denote the
subalgebra of ${\cal A}^{boundary}_{\cal S}$ that leaves invariant
the subspace ${\cal H}^{diffeo}_{y_\alpha ,j_\alpha}$.   All of the
area operators, ${\cal A}[{\cal R}]$  act proportionally to the identity
on these subspaces.  The Wilson loop
operators, $T[\alpha ]$, of loops in the boundary are also all
in these subspaces, as they each commute with the area operators.
But to avoid possible divergences from multiplying Wilson loops,
we may consider the subalgebra of the $T[\alpha ]$'s generated
by the Wilson loops of all 
$\alpha \in {\cal S} - \{ y_\alpha \} $.  It will then be a question of
the representation and boundary conditions how to extend this
to all loops including those that go through the marked points.
Further, for each set of marked points, there exists a subgroup of
$Diff ({\cal S})$ that leaves invariant each 
${\cal H}^{diffeo}_{{\cal S}, y_\alpha , j_\alpha}$, which is
of course given by $Diff({\cal S}- \{ y_\alpha \} )$.  We will then take
each ${\cal A}^{\prime gr}_{{\cal S}, y_\alpha , j_\alpha}$ 
to be generated
by the $T[\alpha ]$ and $D(v)$ for all loops and vector fields
in ${\cal S}-\{ y_\alpha \}$.  These are equal to the full algebras  
${\cal A}^{ gr}_{{\cal S}, y_\alpha , j_\alpha}$ of surface observables
that leave invariant each 
${\cal H}^{diffeo}_{{\cal S}, y_\alpha , j_\alpha}$ up to technical
issues involving the treatement of loops that intersect the
punctures.

We have now come to the point where we can ask about a possible
relationship between quantum gravity and the Chern-Simons 
theories.
For, given any set of marked points and representations of
$SL(2)_q$ on
${\cal S}$ we have two state spaces of interest,
${\cal H}^{CS}_{y_\alpha ,j_\alpha}$ of the corresponding
Chern-Simons theory and ${\cal H}^{diffeo}_{y_\alpha ,j_\alpha}$,
a subspace of the states of quantum gravity.  
We also have
two observable algebras, 
${\cal A}^{CS}_{{\cal S}, y_\alpha , j_\alpha}$
of Chern-Simons theory and 
${\cal A}^{\prime gr}_{{\cal S}, y_\alpha , 
j_\alpha}$,
a subalgebra of observables of general relativity.

We may note that while the states in general relativity are
described so far in terms of ordinary spin networks, for every
$j \leq k/2$ there are representations of both $SU(2)$ and
$SL(2)_q$.  Thus, we have 
this correspondence for each set of $j_\alpha \leq k/2$.

The question of whether there may be any relationship between
these subalgebras and representation spaces depends on the  
choices of dyanamics and boundary conditions 
imposed on quantum gravity.  
Any correspondence must also give
a meaning to the level $k$ of the Chern-Simons theory, 
which is the deformation parameter of the quantum
group, in terms
of the paramters of  quantum gravity.

What is the best possible correspondence that we may hope for
between the states and observables of Chern-Simons theory and
quantum gravity?  The best that may be hoped for is that there
are isomorphisms, 
\f
{\cal A}^{CS}_{{\cal S}, y_\alpha , j_\alpha} =
{\cal A}^{\prime gr}_{{\cal S}, y_\alpha , j_\alpha}  .
\ff
I will show in the next three sections that there is a choice
of boundary conditions that exactly achieves this.  This is
accomplished by requiring that, in the spacetime language,
\f
\left ({1 \over G} e^{AA^\prime} \wedge e^B_{A^\prime}
- {k \over 2\pi} F^{AB}\right )_{\partial {\cal M}} =0 ,
\ff
where $e^{AA^\prime}$ is the frame field one form and 
$F^{AB}$ is the left handed curvature two form, and  
$(...)_{\partial {\cal M}}$ means a form pulled back into the 
boundary.
When expressed in a canonical language, this will become
three conditions on the initial data at each point on the
boundary, $\cal S$,
\f
\left ( \tilde{\Pi}^a_i\epsilon_{abc} 
 - {k \over 2\pi } F_{bc}^i \right )_{\cal S} =0
\ff
where
$\tilde{\Pi}^a_i$ is the momenta conjugate to the conection $A_a^i$. 
In words, we are imposing the condition that the fields, pulled
back to the boundary, are self-dual.  

I will discuss in the next section 
how this condition may be imposed, and
what can be said about how restrictive it is.   

The parameter $k$ will appear as the coefficient of a boundary
term, and, by invariance under large gauge transformations, 
will be required to be an integer, just as in the case of Chern-Simons
theory.  But it will be constrained, by consistency with the
Einstein equations, to be given by,
\f
k={6\pi \over G^2 \Lambda } + \alpha
\ff
where $G$ is Newton's constant, $\Lambda$ is the cosmological 
constant,
and $\alpha$ is a topological $CP$ breaking parameter.

The relations (20) and (21) will later be derived 
from the quantization
of general relativity with the boundary conditions (19).  For now
we proceed to describe the theory that follows, once we are
given these conditions.

The main result of the imposition of these boundary conditions
is a reduction and deformation of the action of each of the 
subalgebras,
${\cal A}^{\prime gr}_{{\cal S},y_\alpha ,j_\alpha }$ 
on the subspaces ${\cal H}^{diffeo}_{y_\alpha , j_\alpha }$ 
so that they are isomorphic to the 
observable algebras ${\cal A}^{CS}_{{\cal S},y_\alpha ,j_\alpha }$
of the corresponding Chern-Simon theories.
This is because, as is straightforward to show, the canonical
form of the boundary condition, (20) when imposed in the
loop representation, 
implies directly the Gauss constraint of Chern-Simons
theory with sources (3), because the $\tilde{\Pi}^{ai}$ integrated
over the boundary, acting on
a state $|\Gamma >^{gr}$ will recieve contributions only at points
where the spin network $\Gamma$ enters the boundary.  We will 
see
the details of this in section VI.

The isomorphism of the observable algebras of the boundaries has
a crucial implication, which is that we must
use quantum spin networks, rather than ordinary spin networks,
to construct the state space of quantum gravity.  
This is because the boundary condition implies, as we have just
said, that the subalgebras  
${\cal A}^{gr}_{{\cal S},y_\alpha ,j_\alpha }$ are isomorphic to the
observer algebras of the corresponding Chern-Simon theories,
and they can only be realized in a space of
states whose basis elements are labeled by quantum spin networks.  
Thus, at least
the portions of the networks in $\Sigma$ that run in
the boundary must be labeled in the restrictive set of
$j\leq k$ that correspond to representations of 
$SL(2)_q$.  

We may ask then if it is necessary that the spin networks remain
$q$-deformed for those parts that travel in the interior
of the space $\Sigma$. Perhaps an edge could be
interpreted as carrying a representation of
$SL(2)_q$ when it meets the boundary, while it carries a
the corresponding representation of $SU(2)$ when it meets
a vertex in the interior.
A simple argument shows that were this to occur, it would
be inconsistent with diffeomorphism invariance.  Let us imagine 
an edge $\gamma$ that runs
from the boundary to a vertex, $v$, where it meets other edges
in the interior.  Let us consider a family of little surfaces
that intersect $\gamma$ at points between the boundary and
the disk.  By diffeomorphism invariance we should expect that
all these surfaces have the same area, for there is no
diffeomorphism invariant structure to distinguish them. Thus,
a surface that approaches arbitrarily closely to the boundary,
${\cal S}$ must have the same area as one that approaches
arbitrarily close to the vertex, $v$.  
However, we know that the area of a surface depends on
$q$ \cite{sethlee-qnet}.  This means that if the edge
$\gamma$ is considered to carry a representation of 
the quantum group as it leaves the surface, it must carry
the same $j$ and $q$ when it meets the vertex; it cannot
somewhere forget the value of $q$.  But this means that 
all lines exiting the vertex must carry representations
of $SL(2)_q$, otherwise the combinatorics of the addition
of quantum spin would not be consistent at the vertex.  One
can then continue this argument to show that every edge in the
interior must be labeled by a representation of 
$SL(2)_q$.

We may ask whether it makes sense to represent the observable
algebra of quantum gravity in terms of quantum spin networks.
This means that the algebra of observables of quantum gravity
accessible on the surface has been deformed, with
$k$ given by (21) being the deformation parameter, in each
subspace ${\cal H}^{diffeo}_{y_\alpha , j_\alpha }$, to the
algebra of observables of the corresponding Chern-Simons
theory.  
We are forced to do this 
by the modifications in the commutation relations
that the boundary conditions impose.  But we may ask if
the deformation in the observable algebra may be extended
consistently to observables on the interior of $\Sigma$, so that
their action on quantum spin networks may be defined.  The answer
is that this can be accomplished by a suitable deformation of the
loop algebra\cite{sethlee-qnet}.  
The result is that all the parameters of 
general relativity, $G, \Lambda $ and $\alpha$ are coded in
the kinematics of the loop algebra.

But once we know that we must use quantum spin networks
as the basis of the states of quantum gravity (in this context)
we have at once, for every set of
punctures and representations, a map from the 
diffeomorphism invariant state space of general relativity
into the direct sum of the physical state spaces of all
the Chern-Simons theories on $\cal S$.  This map is
defined for every $y_\alpha$ and $j_\alpha$,
\f
{\cal N}_{y_\alpha , j_\alpha }: 
{\cal H}^{diffeo}_{y_\alpha ,j_\alpha}
\rightarrow {\cal H}^{CS}_{y_\alpha ,j_\alpha}
\ff
and is given by
\f
{\cal N}_{y_\alpha , j_\alpha }: |\Gamma >^{gr} \rightarrow 
|\Gamma >^{CS} .
\ff
Let me emphasize that this map exists naturally by the axioms of 
topological
quantum field theory and the fact that the isomorphism between the
observable algebras at the boundary requires that quantum spin 
networks
be used in quantum gravity as well.

We may then write
\f
{\cal H}^{diffeo}_{y_\alpha ,j_\alpha}=
{\cal H}^{CS}_{y_\alpha ,j_\alpha} \otimes {\cal K}_{y_\alpha 
,j_\alpha }^{diffeo}
\ff
where ${\cal K}_{y_\alpha , j_\alpha }^{diffeo}$ is 
the kernal of the map
${\cal N}_{y_\alpha , j_\alpha }$.  

Given that the state space ${\cal H}^{CS}_{y_\alpha ,j_\alpha}$ of
the Chern-Simons theory is finite dimensional, we may expect that,
at the level of diffeomorphism invariant states,
these kernals are themselves infinite dimensional.  
In the classical theory, we expect that 
there are an infinite number of diffeomorphism 
invariant observables that
correspond to measurements made at points inside of the boundary,
that must then commute with all the observables defined on the
boundary.   Unfortunately, we do not know many of these observers
explicitly.  One, which is now understood, is the volume
of $\Sigma$.  The corresponding quantum operator has
now been defined, and their eigenstates and eigenvalues
have been characterized\cite{volume}.  
The trivalent spin networks
are precisely eigenstates, while for higher valence networks
the eigenstates are known to lie in the finite dimensional
sectors spanned by the different intertwinings at the vertices.
In both cases the eigenvalues are discrete, and on order of the
Planck scale.  This operator may also be defined, through
the deformation of the loop algebra, to act on the quantum
spin networks\cite{sethlee-qnet}.

Indeed, given the interpretation
of spin network states at the 
kinematical level\cite{volume} 
in which each node carries a certain unit of volume and each
edge a quanta of area, we have a good understanding of what
the different states $|\Gamma >^{diffeo}$ mean, at least
in the limit of large $k$: they are
eigenstates of observables that measure the three geometry
of the interior.  It is clear that, even when the 
geometry of the surface is fixed, which determines the
$y_\alpha $
and $j_\alpha$ of the intersections of $\Gamma$ with the boundary,
there should still be an infinite number of diffeomorphism invariant 
quantum states
of the gravitational field 
in ${\cal H}^{diffeo}_{\cal S}$ for each state 
$|{\cal N}_{y_\alpha ,j_\alpha} \circ \Gamma >^{CS}$ in the
associated Chern-Simons theory.  These correspond to different
spatial diffeomorphism classes of three geometries 
on the interior that, however
agree about the quantum fields that are induced on the boundary.

To show this we need to have more information about the
diffeomorphism invariant observable algebra and its
$q$ deformation.  However,
in the absense of this, we may still attempt to make reasonable
hypotheses about the full diffeomorphism invariant
and physical state spaces.  I would like now to describe two.

\subsection*{Hypothesis I:  Existence of a complete set of
diffeomorphism invariant observables}

Let us consider the diffeomorphism invariant 
configuration space of general
relativity with $\Sigma$ fixed, given the boundary conditions (20),
which I will call ${\cal C}_{diff}$.   It is clear that there
must exist a complete set of coordinates on it
\f
x_{phys}=\{  {\cal A}[{\cal R}], x_{\Sigma} \}
\ff
where $x_{\Sigma}$  label all the diffeomorphism classes
of three metrics on $\Sigma$ that induce the areas 
${\cal A}[{\cal R}]$  on $\cal S$.  This is a weak
assumption,  the 
main issues to be solved to establish it are global issues.
 
If this is the case, then there must be in the full quantum
theory a set of commuting diffeomorphism observables, which I will
call ${\cal O}_\Sigma^I$, where $I$ is a generic
label,  which have the property that an orthonormal
basis of each kernal ${\cal K}_{y_\alpha ,j_\alpha }$ exists  which
are the eigenstates of the $\hat{\cal O}_\Sigma^I$.  I will label
these by $|\lambda^I, y_\alpha ,j_\alpha >$, which are defined
so that
\f
|\lambda^I, y_\alpha ,j_\alpha > \in {\cal K}_{y_\alpha ,j_\alpha }
\ff
and
\f
\hat{\cal O}^I 
|\lambda^I, y_\alpha ,j_\alpha > = \lambda_I 
|\lambda^I, y_\alpha ,j_\alpha > 
\ff
Let $|z>$ be an orthonormal basis of states in 
${\cal H}^{CS}_{y_\alpha , j_\alpha }$ such that, under the inner
product of that theory,
\f
<z |z^\prime >^{CS}_{y_\alpha , j_\alpha } = \delta_{z z^\prime }
\ff
It then follows that there is a basis for diffeomorphism invariant
states of the gravitational field, given by
\f
|y_\alpha , j_\alpha , \lambda_I , z> = |z> \otimes 
|\lambda^I, y_\alpha ,j_\alpha >
\ff
By virtue of the fact that the 
${\cal A}[{\cal R}]$ and ${\cal O}^I $ must be represented by
hermitian observables, these are an orthonormal basis for
the Hilbert space of spatially diffeomorphism and gauge
invariant states.
It then follows that if any two states
$|\Psi >, |\Psi^\prime > \in {\cal H}^{diff}_{y_\alpha , j_\alpha } $
of the form
\f
|\Psi > = |{\cal N} \circ \Psi > \otimes |{\cal K}\Psi >
\ff
where the two factors $|{\cal N} \circ \Psi > $ and 
$|{\cal K}\Psi >$ are in ${\cal H}^{CS}_{y_\alpha , j_\alpha }$
and ${\cal K}_{y_\alpha ,j_\alpha }^{diff}$, respectively, we have
\f
<\Psi |\Psi^\prime>^{diffeo} = 
<{\cal N} \circ \Psi |{\cal N} 
\circ \Psi^\prime >^{CS}_{y_\alpha, j_\alpha }
<{\cal K}\Psi |{\cal K}\Psi^\prime >^{\cal K}_{y_\alpha , j_\alpha } 
\ff
We will see in section VIh. below that this is realized, given
only some rather weak assumptions.

Let us now consider the state
$|\Gamma >^{gr} \in {\cal H}^{diff}_{\cal S} $ associated to the
quantum spin network $\Gamma $.  It then follows from
what we have said, together with the isomorphism (18)
that if we measure an observable $T[\alpha ]$ for a loop
$\alpha \in {\cal S}$ we have 
$|{\cal N}\circ \Gamma > = |\Gamma >^{CS}$
so that,
\f
{<\Gamma |\hat{T}[\alpha ] |\Gamma >^{diffeo} \over 
<\Gamma |\Gamma >^{diffeo} } = 
{<\Gamma |\hat {t}[\alpha ] 
|\Gamma >^{CS} _{y_\alpha^\Gamma , j_\alpha^\Gamma }\over 
<\Gamma |\Gamma >^{CS}_{y_\alpha^\Gamma , j_\alpha^\Gamma } }
\ff
Thus, the expectation value in quantum gravity of an
observable that measures the self-dual curvature at the
surface $\cal S$ is given by an expression in Chern-Simons
theory.  Furthermore, this expression may be evaluated in
closed form; it is proportational to the Kauffman bracket
of the knotted graph $\Gamma \circ \Gamma^* \cup \alpha $.

\subsection*{ Hypothesis II:  The kernals are trivial in
the physical state space}

The kernals ${\cal K}_{y_\alpha ,j_\alpha }^{diffeo}$ contain precisely
the information about the diffeomorphism invariant 
state of the gravitational
field that is not accessible to the observers who can only measure
the system by measuring the metric and connection 
on the surface $\cal S$.  It is clear from the construction we
have just gone through that in the case we are considering
diffeomorphism invariant states there is an infinite
amount of such information. 
However, what is the situation with respect to physical
states, those that are solutions to the Hamiltonian constraint,
and are therefor constrained by the dynamics of the theory?

As I will show in section V, with the boundary 
conditions (20) the
hamiltonian constraint must vanish on the boundary, so we
know that the observables ${\cal A}_{\cal S}^{boundary}$
are all physical observables.  We also know, from the action
of the Hamiltonian constraint, in any of the forms given 
in \cite{lp1,ham1,ham2,jorgerodolfo} that the individual 
spin network states $|\Gamma >$
are not physical states.  However, because the areas on the
boundary commute with the hamiltonian constraint, the action
will not mix states in different subspaces
${\cal H}^{gr}_{y_\alpha ,j_\alpha}$.

Thus, we know that the physical state space, 
${\cal H}^{phys}_\Sigma$,
defined as the subspace of ${\cal H}^{phys}_\Sigma$
that is annihilated by the hamiltonian constraint, with
lapse vanishing at the boundary, is given by the 
same decomposition
\f
{\cal H}^{phys}_{\Sigma}= \sum_{n=0}^\infty 
\sum_{j_1...j_n }\int d^2y_1...d^2y_n
{\cal H}^{phys}_{y_\alpha ,j_\alpha}
\ff
where each
\f
{\cal H}^{phys}_{y_\alpha ,j_\alpha}=
{\cal H}^{CS}_{y_\alpha , j_\alpha }
\otimes
{\cal K}_{y_\alpha ,j_\alpha }^{phys}
\ff
It then follows that the physical kernals 
${\cal K}_{y_\alpha ,j_\alpha }^{phys}$ are each subspaces
of the diffeomorphism invariant kernals
${\cal K}_{y_\alpha ,j_\alpha }$.  

This leads to the important result that the expression (32) must
hold also for the physical expectation values of the physical
observables $T[\alpha ]$ on physical states, as long as 
$\alpha \in {\cal S}$.  We are similarly able to evaluate the
expectation value of any of the physical area observables.
This means that we are able to evaluate the expectation values of
an infinite and nontrivial set of physical observables,
which characterize observations we may make on
the boundary,  in closed form, even in the absense of further
information about the physical states.

To say more than this, we need information about the
physical kernals, ${\cal K}_{y_\alpha ,j_\alpha }^{phys}$.
In the light of what we have just said
I would like to propose that we consider the following
simple conjecture, which is that these physical kernals are
all trivial.  This is equivalent to the physical hypothesis
that all information about a physical state of the
gravitational field in a region $\Sigma$ surrounded by
a finite boundary is accessible by measuring the metric
and connection induced in that boundary.  Mathematically,
it is equivalent to the conjecture the the physical state
space is precisely,
\f
{\cal H}^{phys}_{\Sigma}= \sum_{n=0}^\infty 
\sum_{j_1...j_n }\int d^2y_1...d^2y_n
{\cal H}^{CS}_{y_\alpha ,j_\alpha}
\ff

In this case, the physical quantum states must be exactly
the states $|\Gamma , \Sigma >^{CS}$ defined
by the Chern-Simons path integral, but now with appropriate
boundary conditions.  These may be considered to be the
generalizations of the Bruegmann-Gambini-Pullin states\cite{BGP}
to the case of the  boundary conditions (20).   If this
conjecture is correct, then we come to the conclusion that
the number of solutions to the Hamiltonian constraint that
match the boundary conditions for each set of
colored punctures, $(y_\alpha ,j_\alpha )$ is equal only to the 
number
of states of the associated Chern-Simons theory.   All other
states, such as the infinite numbers of states known which
are associated with knot 
classes\cite{lp1,spinnet-us}, must fail to match
the boundary conditions on the surfaces.  Another way
to say this is that, as proposed by Crane\cite{louis}, all 
physical quantum states
found by performing the loop  transform of the Kodama state 
of the connection representation,
\f
\psi_{Kodama} [A]  = e^{{ik \over 4 \pi} \int_{\Sigma} Y_{CS} (A) }
\ff
with the differences in the states 
being due to boundary conditions imposed on the transform.

I would like to give several arguments for this proposal:

1) As already mentioned, it is a reasonable hypotheses
that all physical
information about the interior of a finite region should be
determinable by measurements made on its boundary.   

I may note that even if this is
not the case, then the description in terms of (35) should apply
at least to those cases in which, because of the presence of
a horizon, we can only measure
observables on the boundary $\cal S$.  If we take an operational
point of view, in which the Hilbert space is to be spanned by the
eigenstates of a complete set of commuting operators corresponding
to observables that we can in fact measure, then there is no sense
to introducing extra factors in the Hilbert space that are 
distinguished only
by operators that cannot be measured.

2)  This does not conflict with the usual practice in conventional
quantum field theory.  There it is usually established that the
physical Hilbert space may be spanned by 
scattering states that can
all be distinguished by measurements made by observables far from 
the region in which interactions take place.  
In a diffeomorphism invariant system, there is no meaning, independent
of a given state,
to being far from the interacting region.  We may try to 
investigate the notion
of asymptotic observables, which has a diffeomorphism invariant
meaning, but there are special problems associated with this. At least
in the case of spatial infinity, it is doubtful that there are more
than a finite number of observables defined at spatial infinity.

The alternative is to fix a finite boundary, and define the physically
meaningful states by what an observer on the boundary could 
measure.
This seems to be the most useful diffeomorphism invariant extension 
of
the usual practice of defining the physical Hilbert space in terms of
asymptotic states, and it leads, in the case under
consideration, to the conclusion that the kernals 
${\cal K}_{y_\alpha ,j_\alpha }^{phys}$ must be trivial.

To get consistency with the usual practice, what is required is
only that the dimensionality of the state space associated with a
fixed two sphere metric on the boundary go to infinity as the
area of the boundary itself is taken to infinity.  This proposal 
satisfies
that requirement, as we will see explicitly.

3)  This does not mean that the state space of quantum gravity is
finite dimensional.  It only implies that the degenerate
subspaces of states that agree about all measurements of areas of 
regions
of the boundary are finite dimensional.  The
whole physical state space is infinite dimensional because it is
composed of sums and integrals over finite factors according to (35).
What this analysis has given us is a complete description of each
of the factors as the state space of a Chern-Simons theory.

4)  It may still seeem that the state space given by (35) is
too small to represent quantum mechanically the degrees of
freedom of the gravitational field in the interior of $\Sigma$,
as there are, according to the classical canonical analysis,
two degrees of freedom per point.  However, there is the old
argument, given first by Beckenstein\cite{beckenstein}, and 
developed by Hawking\cite{hawking},
and more recently 
by 't Hooft\cite{gerard-holographic} and 
Susskind\cite{lenny-lorentz}, 
that this drastically
overcounts the actual degrees of freedom in any region surrounded
by a surface of finite area.  The reason is that,
due to the phenomena of gravitational collapse, almost all of
the configurations that one would take into account by counting
two degrees per point of volume would be surrounded by an
event horizon.  Instead, Beckenstein conjectured that the amount
of information that can be stored in a region surrounded by
a boundary $\cal S$ of finite area ${\cal A}[{\cal S}]$ is finite
and is bounded by,
\f
I_{\cal S} = c  {{\cal A}[{\cal S}] \over l_{Pl}^2}
\ff
where $c$ is a fixed constant.  This means
that the dimension of the physical state space for the corresponding
quantum system must be finite and bounded by
\f
dim \  {\cal H}^{phy}_{\cal S} \leq e^{I_{\cal S}LN(2)}
\ff
Motivated by this, 't Hooft and Susskind have made the
``holographic hypothesis", which is that the states
in ${\cal H}^{phy}_{\cal S}$ can be actually represented in terms
of a finite quantum field theory on $\cal S$ 
\cite{gerard-holographic,lenny-lorentz}.

The hypothesis (35) for the physical state space may be considered
to be a realization of the holographic hypothesis.  We may further
show that it yields also the Beckenstein bound, at
least in the case of small cosmological constant.

To see this,
let us assume that we have measured the metric on the spatial
boundary as accurately as we can, given the discrete nature of
the quantum geometry.
How much information yet remains that could code a description
of the quantum state inside the boundary?  

The most precise possible measurement of the metric geometry of
$\cal S$ reduces us to a single subspace of the physical state
space, 
${\cal H}^{CS}_{{\cal S},y_\alpha ,j_\alpha} $.  The information
that is then given by specifying the exact state within this
subspace is 
\f
I_{{\cal S},y_\alpha ,j_\alpha}= LNdim
\left ({\cal H}^{CS}_{{\cal S},y_\alpha ,j_\alpha} \right )
= \sum_\alpha LN (2j_\alpha + 1)
\ff
At the same time the area is given in the large $k$ limit by (16).
(We may note that for finite $k$ there are corrections to
this area formula\cite{sethlee-qnet}, but 
they are irrelevant in this case.)

To proceed further we now want to maximize both the amount
of information contained in any large but finite region of $\cal S$ 
and the accuracy with which the
quantum geometry approximate a classical metric.  As both
the area and the information (at large $k$) are additive quanties,
each is maximized in the same way, which is when werepresent
the two metric by the maximum number of punctures, each of
which has the smallest possible spin and so 
contributes  the minimal possible area.  To see this note
that, if all punctures have the same spin,  $j$,
$I/A \approx 2 LN(2j+1)/\sqrt{j(j+1)}$ which is maximized
for the smallest value $j=1/2$.  We then have
\f
I_{\cal S} = {4 (LN(2))^2 \over \sqrt{3} } \ 
{{\cal A}[{\cal S}] \over l_{Pl}^2}
\ff

Thus, given the conjecture that all the information about the
quantum state in $\Sigma$ can be gotten by measuring the metric
and self-dual connection induced on the surface $\cal S$, we
reproduce the Beckenstein bound that the maximum amount
of information that can be contained within a surface is 
equal to a fixed constant times its area, once the metric of the
surface is measured as accurately as possible.

We may invert this argument to conclude that, if the physical
kernals ${\cal K}^{phys}_{y_\alpha ,j_\alpha }$ are non-trivial,
the Beckenstein bound will  be violated (unless the 
dimensionality of the kernals also grows like 
$exp({\cal A}[{\cal S}])$.  But this would
make it impossible to give a statistical interpretation
to black hole entropy.   

5)  In the particular cases that the spacetime boundary
$\partial {\cal M}$ is the horizon for a family of observers,
there is an argument, due to Jacobson\cite{ted-new}, 
which derives the Einstein
equations from the following short list
of assumptions: 1) the laws of thermodynamics, 
2) standard special
relativistic field theory holds in small regions for inertial
observers 3) the information 
not measurable by the observers, because of the existence of
the horizon, is proportional to its area.

This would suggest that there can be no more information in the
quantum state of the gravitational field in the interior of
$\cal M$ than is contained in the hypothesized physical
state space (35).

\section*{IV. The classical theory with self-dual  boundary conditions}

All the conclusions of the previous section were based on the
assumption that a quantum theory can be defined
to describe the quantum physics of the gravitational 
field in the interior of a surface $\cal S$ in such a way that three
conditions are satisfied:

1)  The self-dual boundary condition (20) is imposed.

2)  The gauge transformations generated by the hamiltonian
and diffeomorphism constraints leave the boundary $\cal S$
fixed.

3)  The internal gauge transformations act on the boundary.

In this section we will derive these conditions by imposing the
condition (19) that the spacetime curvature be self-dual, with
a cosmological constant, at the surface $\cal S$.
 
The most direct way  to do 
this is to use the 
Capovilla-Dell-Jacobson forms of the action\cite{CDJ}, which
is given in terms of the left handed spacetime connection one form,
$A^i$, and
a matrix of scalar fields, $\Phi_{ij}$, which is
restricted to satisfy
\f
\Phi_{ij}=\Phi_{ji} , \ \ \ \mbox{and} \sum_i \Phi_{ii} = \lambda .
\ff
where $\lambda=G^2 \Lambda$ is the 
cosmological constant in Planck units.  We will see that it
will be essential here to consider the case in which $\lambda$ is
not zero.
Here $i,j,k$ are the $SU(2)_{Left}$
indices.  To write the action we begin with the $CDJ$ form, which is,
\f
S_{CDJ}= {1 \over 2} \int_{\cal M} F^i \wedge F^j (\Phi^{-1})_{ij}
\ff

This action does not by itself give a good variational principle in
the case that spacetime has boundaries.  
To define a consistent variational
principle we must choose boundary conditions at $\partial {\cal M}$
and add a boundary term to the action such that the variational
derivative of the total action exists.  One way to do this,
which leads to the self-dual boundary conditions (20) is 
to use the
fact that at points of spacetime at which $\Phi$ is diagonal
the  Weyl tensor must be self-dual\cite{chopin-thesis}.  
Roughly speaking, this corresponds to solutions
in which there are only left-handed gravitational waves at the 
boundary.

We thus proceed by imposing the boundary condition that
\f
\Phi_{ij}|_{\partial {\cal M}} = { \lambda \over 3} \delta_{ij}
\ff
We then must add a boundary term to the action (42) 
whose variation
will cancell the boundary term in its variation.  Interestingly
enough, the correct boundary term is the Chern-Simons action of the
remaining, left-handed part of the curvature 
on the boundary\cite{ls-review}.

Thus, the full action principle we will be interested in is,
\f
S= {1\over 2} \int_{\cal M} F^i \wedge F^j (\Phi^{-1})_{ij}- 
{3 \over 2 \lambda} \int_{\partial {\cal M}} Y_{CS} (A)
\ff
where the Chern-Simons form is given by,
\f
Y_{CS}(A) = A^i \wedge dA^i + 
{ 1\over 3} \epsilon_{ijk}A^i \wedge A^j \wedge A^k
\ff
As the reader may check, with the boundary condition (43)
the variation of this action is a pure volume term,
giving the field equations
\f
{\delta S \over \delta A_{\alpha^i} } = 
- \epsilon^{\alpha \beta \gamma \delta}
{\cal D}_{\beta} \left ( F_{\gamma \delta}^j (\Phi^{-1})_{ij} \right ) =0
\ff
and
\f
{\delta S \over \delta \phi_{ij} }= F^k \wedge F^l
[(\Phi^{-1})_{ki}(\Phi^{-1})_{lj}-{1 \over 3} \delta_{kl}
(\Phi^{-1})_{mn}(\Phi^{-1})^{mn}
=0 .
\ff
We may also add a topological term to the action (44) to get,
\f
S= {1\over 2} \int_{\cal M} F^i \wedge F^j (\Phi^{-1})_{ij}
+ {\alpha \over 4 \pi} \int_{\cal M} F^i \wedge F^i  - 
\left ( {3 \over 2 \lambda} +{\alpha \over 4 \pi} \right ) 
\int_{\partial {\cal M}} Y_{CS} (A)
\ff
This makes the theory $CP$ violating, as $\alpha$ plays the same
role as the strong $CP$ breaking parameters 
in $QCD$\cite{ABJ}.

To complete the definition of the theory, we must  note that
gauge invariance imposes an additional restriction on our theory.
As is well known, the Chern-Simons action is not invariant under
large gauge transformations, instead it transforms as
$\int Y_{CS} \rightarrow \int Y_{CS} + 8\pi^2$.  If we require that the
Minkowskian path integral $e^{iS}$ is to be invariant under
such transformations, this imposes the requirement that
\f
k = {6\pi \over \lambda} +\alpha
\ff
where $k$ is an integer, and in fact will be the {\it level} of the
Chern-Simons theory.  

Thus, gauge invariance implies a quantization of the cosmological 
constant.  Significantly, we see that small cosmological constant
corresponds to large $k$, which is the semiclassical limit of the
Chern-Simons theory.

One may be suspicious that these results are somehow an artifact
of the Capovilla-Dell-Jacobson formalism.  However, it is easy to see
that one gets the same results using the self-dual actions
given earlier in \cite{tedmesam}.  In the presense of the finite 
boundary,
that action must be extended by a boundary term so that 
the variation is a total derivative, so we have
\f
S_{SD}=\int_{\cal M} \left ( {1 \over G} 
e^{AA^\prime}\wedge e^{B}_{B^\prime}
\wedge F_{AB}(A) +{\alpha \over 2 \pi } F^{AB}\wedge F_{AB}
+ \Lambda det(E) \right )
- {k \over 2\pi} \int_{\partial {\cal M}} Y_{CS} (A)
\ff
When we vary this we find,
\begin{eqnarray}
\delta S_{SD} &=& \int_{\cal M} \left ( {1 \over G} 
\delta A_{AB}  \wedge
{\cal D} \wedge 
(  e^{AA^\prime}\wedge e^{B}_{B^\prime} )
+ \delta e^{AA^\prime} \wedge ({2 \over G} e^B_{A^\prime}
\wedge F_{AB} + 
3 \Lambda e_A^{B^\prime} 
\wedge e_{B^\prime C} \wedge e^C_{A^\prime}  \right ) 
\nonumber \\
&+& \int_{\partial {\cal M}} \delta_{AB} \wedge
\left [ {1 \over G} e^{AA^\prime}\wedge e^{B}_{B^\prime}
- {(k-\alpha )  \over 2 \pi} F_{AB}
\right ]
\end{eqnarray}

For the equations of motion to be well defined, the surface
term must vanish.  We see from this expression that one way
to accomplish this is to fix $A_{AB}$ on the boundary so that
its variation vanishes there.  We may note that if
$k-\alpha$ vanishes, this is the only alternative, as long
as we want the metric on the boundary to be non-degenerate.
However,  there is another way to cancel the surface term
which is to choice the self-dual condition (19) on the boundary,
so that
\f
{1 \over G} e^{AA^\prime}\wedge e^{B}_{B^\prime}
- {(k-\alpha )  \over 2 \pi} F_{AB}=0
\ff
This is an acceptable boundary condition, as long as it is
consistent with the field equations at the boundary.  This
is necessary because this allows both the connection and frame
field to vary at the boundary, so that the field equations
must be satisfied there.  If we take $e^{B}_{A^\prime}\wedge$
times (52), we see that this is equal to the frame field equation
at the boundary, so long as the condition (49) on the constants
is imposed.   Thus, we see that an examination of the variational
principle in this form leads to the same conclusons as the 
variational principle according to the CDJ formalism.

Before proceeding, we must say a word about the effect of
imposing these boundaries on the space of solutions.  It is
important to note that the condition we have imposed, which
is that the pull back of the curvature to the boundary satisfies
the self-dual condition (52) is significantly weaker than a
requirement that the whole curvature tensor, evaluated at
the boundary, is self-dual.  If self-duality is imposed on all
components of the tensor, then it may be argued that, at least
in the analytic case, the field is self-dual in the whole manifold.
This may be shown directly by writing down the Bianchi
equations.  The conditions that are being imposed here are weaker
than that, in terms of initial data they involve, as we will
see in the next section,  imposing  the constraint (20), which
involves three of the nine components of the self-dual equations.
A linearized analysis of these conditions may be performed that 
shows that, at least locally, we may expect an infinite number of
non-self-dual solutions to Einstein's equations with these 
boundary conditions, this will be discussed elsewhere.

We may note also that were the whole self-dual conditions being
imposed in the Minkowskian case, then there would  be
only one solution, as the whole solution would be self-dual, but there
is only self-dual solution of Minkowskian signature, which is
DeSitter spacetime.  How these weaker conditions interact with the
Minkowskian reality conditions is an important open question.
For this reason, and because the Chern-Simon theory is simplest
in the compact case, the considerations of this paper are restricted
to Euclidean signature.

We now proceed to the Hamiltonian analysis, based on the
CDJ form of the action.

\section*{V. Hamiltonian dynamics, including the 
boundary conditions}

We now proceed to construct the hamiltonian dynamics that
corresponds to the variational principle (44).  To keep
the analysis simple, we will restrict ourselves to the
case that the $CP$ violating phase $\alpha$ is set
to zero.  When it is included, the hamiltonian analysis
is a bit more complicated, following the lines described
in \cite{ABJ}.    

To proceed, we will 
assume a  $3+1$ splitting, so that spatial indices,
$a,b,c$ will correspond to coordinates in $\Sigma$, while
$t$ will be a coordinate on $R$.   We then find the canonical
momenta has a boundary contribution, and is of the form
\f
\tilde{\Pi}^{ai}(x) \equiv {\delta S \over \delta \dot{A}_a^i (x)}
= {1 \over 2} \epsilon^{abc} F_{bc}^j (\Phi^{-1})_{ij}(x)
- {3 \over 2 \lambda} \int_{\partial \Sigma } d^2S^{ab}(\sigma)
A_b^i (S(\sigma )) \delta^3(x, S(\sigma ))
\ff
Here $\sigma$ are the two coordinates on $\partial \Sigma$, which
we denote here as $S$.

As much of what follows depends on the fact of there being a 
contribution
to the canonical momenta from the boundary, it is important to note
why this must be there.  The boundary term is there to cancell an
integration by parts when one takes the variation of the action by
the left handed connection, $A^i$.  However, the canonical momenta
is the variation of the action by a derivative of $A^i$, hence the 
cancellation
does not hold in this case and one picks up a boundary contribution.

\subsection*{Va.  The second class 
constraints associated with the boundary}

We must first take care of
the fact that   the presence of the  boundary term leads to primary
constraints associated with points of the boundary.  
We may recall that this is 
what happens in the actual Chern-Simons theory,
when the momentum is given only by the boundary term, so that
$\pi_{CS}^{\alpha i} = \epsilon^{\alpha \beta } A^i_\beta$, where
$\alpha$ and $\beta$ are coordinates in the two surface.  However
there is an important difference, due to the fact that for Chern-
Simons
theory space {\it is} the two dimensional boundary, whereas in the
present case it is the three dimensional space $\Sigma$.  As the
momenta are three dimensional densities, if there are constraints 
that
arise from their definition, they must be defined by integrating the
definition of the momenta against smooth functions on $\Sigma$.
If we smear the definition (53) against smooth 
functions $f_{ai}(x)$ we
find a set of relations,
\f
J(f) \equiv \int_\Sigma f_{ai}( \tilde{\Pi}^{ai}-\tilde{B}^{aj}
(\Phi^{-1})_{ij})
-\int_{\partial \Sigma}d^2S^{ab}f_{ai}(S(\sigma))A_b^i(S(\sigma))
=0
\ff
where we have used the useful notation,
$\tilde{B}^{ai}= {1\over 2} \epsilon^{abc}F_{bc}^i$.

These $J(f)$ are primary constraints, that have a second class
algebra due to the presense of the boundary. We are now
about to invert those constraints, to find the Dirac brackets,
which we will then use to construct the quantum theory.
What follows is a bit of a technical exercise, those uninterested
in the details may skip to the end of this section, where the
resulting modification of the loop algebra is displayed in 
eq. (66).  The result, as we will see, is quite intuitive: loop
observables obey the algera of observables of quantum gravity
for those portions of the loops in the interior of the manifold
$\Sigma$, while they satisfy the algebra (7) of Chern-Simons 
theory
for those portions of the loops that travel in the boundary
$\cal S$.
 
To define the Poisson algebra of these constraints, we must define
 the brackets of the fields $A$ and $\Pi^{ai}$ with the 
$\Phi$.   We would
like to do this in such a way that the conventional Poisson structure
is returned in the absence of the boundary, or, equivalently, for
all one forms $f_{ai}$ whose pull back to the boundary vanishes.
This means we must take
\f
\{ \tilde{\Pi}^{ai}(x), \tilde{E}^{ai}(y) \} =0
\ff
where 
\f
\tilde{E}^{ai}(x) \equiv \tilde{B}^{aj}(\Phi^{-1})_{ij},
\ff
 since in the
absense of the boundary contribution 
$\tilde{\Pi}^{ai}(x)=\tilde{E}^{aj} $.
Using this we find the Poisson algebra,
\f
\{ J (f) , J(g) \} = {3 \over \lambda} \int_{\partial \Sigma } d^2S^{ab}
(\sigma )
f_{ai}(S(\sigma )) g_{b}^i(S(\sigma ))
\ff

Thus, whenever the pull back of the 
one forms $f_{ai}$ and $g_{bi}$ to
the boundary are nonvanishing, we have second class constraints.
To define the Poisson structure of the theory we must invert these
to construct Dirac brackets.  To invert these brackets we must 
construct an appropriate regularization of this  second class algebra,
and then define the inversion through a suitable limit.  We may do 
this
in the following way.  First let us define an appropriate set of 
smearing
fuctions, associated to ribbons in $\Sigma$.  These ribbons are one
parameter families of curves, $\alpha_{\tau}(s)$, where 
$0 \leq \tau \leq 1 $ labels the curves, each of which is
parameterized by $0 \leq s \leq 1$.  These curves then define a 
surface
in $\Sigma$, which may or may not intersect the boundary.  This 
construction mimics the strip regularization of the loop algebra 
\cite{ls-review}.
We then define smearing functions.
\f
g^{\alpha_\tau}_{ai} (x) \equiv \int ds \int d\tau \epsilon_{abc}
\dot{\alpha}^b_\tau \alpha^{\prime c}_\tau \delta^3 (x, 
\alpha_\tau(s))
e_i^\alpha
\ff 
where $\alpha^{\prime c}_\tau = \partial \alpha^{c}_\tau /\partial 
\tau $
and $e^g_i$ is a Lie algebra element.  It is then a simple calculation to
show that if $J_{\alpha_\tau}= J(g_{\alpha_\tau} )$,
\f
\{ J_{\alpha_\tau} , J_{\beta_\tau^\prime } \} =
-{ 3 \over \lambda } Int ( \tilde{\alpha} , \tilde{\beta } ) 
(e^\alpha_i e^{\beta i} )
\ff
where $\tilde{\alpha}$ is the loop which is the intersection of the
ribbon with the boundary $\partial \Sigma $ (if there is no 
intersection
it is the zero loop).  We may note that as the ribbon must be in 
$\Sigma$,
its intersection of the boundary must consist of some segments of 
loops.

To invert these relations we are going to represent this algebra as 
the
limit of a sequence of finite algebras, each associated with a lattice
associated to the surface.  To do this we impose a set of coordinates
$\sigma^{\hat{\alpha}}$ on the boundary, and using this construct
a lattice of points with spacing $L$.  The points of the lattice will
be denoted $\hat{n}^L$, and to each there are two line segments,
denoted $\gamma_{\hat{n}\hat{\alpha}}^L$ that each extend in the 
$\hat{\alpha}$
direction for a distance $L/2$ to each side of the point $\hat{n}$.
These lines then form a family of crosses centered on the points 
$\hat{n}^L$.
To each line we construct a ribbon 
$\gamma_{\hat{n}\hat{\alpha}\tau}^L$
such that 
$\gamma_{\hat{n}\hat{\alpha}\tau=0}^L=
\gamma_{\hat{n}\hat{\alpha}}^L$.
We also impose the requirement that as we take $L\rightarrow 0$ 
the
whole ribbon approaches the boundary, so that 
\f
\lim_{L \rightarrow 0}  \gamma_{\hat{n}\hat{\alpha}\tau}^L \in 
\partial \Sigma 
\ff
for all $\tau$.  We then define constraints associated to each of
the lines in the lattice by
\f
J_{\hat{n} \hat{\alpha}}^{\hat{i} L} = J_{
\gamma_{\hat{n}\hat{\alpha}\tau}^L}|_{e^i = \delta_{i\hat{i}}}
\ff
whose algebra is given by
\f
\{ J_{\hat{n} \hat{\alpha}}^{\hat{i} L} , J_{\hat{m} 
\hat{\beta}}^{\hat{j}L}   \} 
= - {3 \over \lambda } 
\delta_{\hat{n}\hat{m}}\epsilon_{\hat{a}\hat{b}}
\delta^{\hat{i}\hat{j} } .
\ff
It is clear that this constitutes a regularization of the second class 
algebra.
That is, 
given a one form $g_{\alpha i}$ on $\partial \Sigma$, one
can easily show that,
\f
\{ J(g), J(g^\prime ) \} = \lim_{L \rightarrow 0 } 
\sum_{\hat{n}\hat{\alpha}}
\sum_{\hat{m}\hat{\beta}} g(\hat{n})_{\hat{\alpha}}^{\hat{i}}
g(\hat{m})_{\hat{\beta}}^{\hat{j} \prime} 
\{ J_{\hat{n} \hat{\alpha}}^{\hat{i} L} , J_{\hat{m} 
\hat{\beta}}^{\hat{j} L} \}
\ff
We may then define the Dirac brackets through the corresponding 
limit
\f
\{ A, B \}_D = \lim_{L \rightarrow 0} \{ A, B\}_{D,L}
\ff
where the regulated Dirac bracket is
\f
\{ A,B \}_{D,L} \equiv \{ A,B \} -{\lambda \over 3} \sum_{\hat{n}}
\epsilon^{\hat{\alpha}\hat{\beta}} 
\{ A, J_{\hat{n} \hat{\alpha}}^{\hat{i} L}  \}
\{ J_{\hat{n} \hat{\beta }}^{\hat{i} L} , B \} 
\ff

Using these relations we can check that the loop algebra is modified
in the cases that the loops have segments that run in the boundary.
The basic relation, which may be derived from (64), is that
\f
\{ T[\alpha ] , T[\beta ] \}_D = -{ \lambda \over 3 } 
\sum_{\tilde{\alpha} \cup \tilde{\beta} }   
Int[ \tilde{\alpha},
\tilde{\beta} ]^{\cal S} \left (
T[\alpha \circ \beta ] - T[\alpha \circ \beta^{-1} ] 
\right )
\ff
where, here $\tilde{\alpha}$ is the intersection of the curve with the
boundary, $Int[]^{\cal S}$ is the intersection 
number in the two 
dimensional
boundary and the sum is over the set of 
points $\tilde{\alpha} \cup \tilde{\beta} $ 
where the curves intersect
in the boundary.

But, (66) is exactly the loop algebra of Chern-Simons theory, with the
relation (49), with the $CP$ breakiing parameter
$\alpha =0$.  Thus, we 
see that when the loops run in the boundary, 
the
loop algebra is deformed by terms that come from Chern-Simons 
theory.

For loops that do not intersect the boundary, the second class 
constraints
have no effect and the usual loop algebra is satisfied.
 
Having taken care of the second class constraints
associated with the boundary, we may go on with the construction of
the constrained Hamiltonian dynamics of our theory.

\subsection*{Vb.  The Gauss's law constraint, in the presence 
of the boundary}

We next turn to consideration of the first class 
constraints in the theory.  
First of
all the momenta conjugate to $A_0^i$ vanish, from which we find
as secondary constraints Gauss's law,
\f
\tilde{\cal G}^i(x) \equiv {\delta S \over \delta A_0^i (x)}
= {\cal D}_a \left ( \tilde{B}^{aj}(\Phi^{-1})_{ij} \right )
\ff
Because of the cancellation coming from the boundary condition this,
like all equations of motion, is local in $\Sigma$.  However, when we
express it in terms of the canonical commenta (53) 
we find a boundary
piece.  To see this it is easiest to consider the constraint smeared
with smooth functions $\Lambda^i$ on $\Sigma$,
\begin{eqnarray}
{\cal G}(\Lambda ) &\equiv & \int_\Sigma \tilde{\cal G}^i \Lambda^i
=0
\\
&=& \int_\Sigma d^3x \Lambda^i {\cal D}_a \tilde{\Pi}^{ai}- 
{3 \over 2 \lambda } \int_{\partial \Sigma} d^2S^{ab} (\sigma )
F_{ab}^i (S(\sigma )) \Lambda^i (S(\sigma ))
\nonumber \\
&=& 
-\int_\Sigma d^3x ({\cal D}_a \Lambda^i )  \tilde{\Pi}^{ai}
- {1\over 2} 
\int_{\partial \Sigma} d^2S^{ab} (\sigma ) \Lambda^i (S(\sigma ))
\left ( \tilde{\Pi}^c\epsilon_{abc} +
{3 \over \lambda } F_{ab}^i (S(\sigma )) 
\right ) 
\nonumber
\end{eqnarray}

We may note that the surface term is precisely the self-dual 
constraint
(20) with $k=6\pi/\lambda$.

It is interesting to look at the variation of our fields generated
by this constraint.  We have,
\begin{eqnarray}
\delta_\Lambda \int_\Sigma \Pi^{ai} f_{ai} &\equiv&
\{ {\cal G}(\Lambda ) , \int_\Sigma \Pi^{ai} f_{ai} \}
\\
&=& \int_\Sigma \Lambda^i \epsilon_{ijk} f_a^j \Pi^{ak}
-{3 \over 2 \lambda} \int_{\partial \Sigma} d^2S^{ab} (\sigma )
({\cal D}_b \Lambda^i (S(\sigma ) ) f_{ai}(S(\sigma ))
\nonumber \\
\delta_\Lambda \int_\Sigma g^{ai} A_{ai}  &\equiv&
\{ {\cal G}(\Lambda ) , \int_\Sigma g^{ai} A_{ai} \}
= \int ({\cal D}_a \Lambda_i ) g^{ai}
\end{eqnarray}

This tells us that, in spite of the boundary term in the definition
of ${\cal G}(\Lambda )$, for all smooth $\Lambda$ it generates
a motion on the phase space.  The specific form of the variation
of the momenta is also interesting; it tells us that restricted
to the boundary the canonical momenta transforms like 
a connection.
This is exactly what must happen, because, by virtue of the 
second class constraints associated with the 
boundary, restricted to
the boundary the momenta is the connection, as is the case
in Chern-Simons theory.

Further, we may note that  for
arbitrary $\Lambda$ the algebra of the 
${\cal G}(\Lambda )$ is still first class.  It is
easily established that the additional term does not 
change the algebra,
so that 
\f
\{ {\cal G}(\Lambda ) , {\cal G}(\Lambda^\prime ) \} = 2
{\cal G}(\Lambda \times \Lambda^\prime ) ,
\ff
where the $\times$
is the $SU(2)$ product.  This occurs because a possible singular term
coming from the Poisson bracket of the two boundary terms 
vanishes by virtue of a symmetry.  

Thus, we take as the set of Gauss's law constraints all 
${\cal G}(\Lambda )$ including those in which $\Lambda^i$ is
nonvanishing on the boundary.

\subsection*{Vc.  The diffeomorphism and hamiltonian constraints}

We may now go on to discuss the diffeomporphism 
and hamiltonian
constraints of the theory.  These arise as primary 
constraints that are
imosed by the restrictions (41) on the form of $\Phi_{ij}$.  
Because
$\Phi_{ij}$ has been fixed on the boundary, these constraints exist
for every open set in the interior of $\Sigma$.   
Thus, we may note that for all smooth
$v^a$ on $\Sigma$ the following identity holds,
\f
0 = \int_\Sigma \epsilon_{abc}v^a 
\tilde{B}^{bi}\tilde{B}^{cj}
(\Phi^{-1})_{ij}
\ff
as a result only of the symmetry of the $\Phi_{ij}$.  If we use the
defining relation of the momenta, (53), we get a 
primary constraint,
which may be written as,
\f
0= {\cal D}(v)= \int_\Sigma v^a F_{ab}^i\tilde{\Pi}^{bi} + {3 \over 
\lambda}
\int_{\partial \Sigma} d^2S^{bc}v^aF_{ab}^i A_c^i
\ff
However, this does not generate a canonical transformation for all
$v^a$ on $\Sigma$ because, as may be easily checked, it is not
functionally differentiable unless conditions 
are imposed on the $v^a$.
These are that
\f
v^a|_{\partial \Sigma} =0 \ \ \ \mbox{and} \ \ 
\partial_r v^r|_{\partial \Sigma} =0
\ff
where the $r$ is any coordinate direction not in the two surface.  

However, it might be the case, as it is for the Gauss law constraints,
that the ${\cal D}(v)$ still form a closed algebra for all $v^a$, 
including
those that do not vanish on the boundary.  A simple check shows 
that
this is not because, 
while the Poisson bracket of two ${\cal D}(v)$'s is defined,
even when smeared with vector fields  
that don't vanish on the boundary, the algebra does
not close unless the vector fields do vanish there.   
Instead, we have,
\f
\{ {\cal D}(v) , {\cal D}(w) \} = \int_\Sigma 
({\cal L}_v w^a )F_{ab}^i\tilde{\Pi}^{bi} + {6 \over \lambda}
\int_{\partial \Sigma} d^2S^{bc}F_{ab}^i F_{dc}^i v^a w^c
\ff
Thus,  there is no meaningful sense in which the diffeomorphism
group can be extended beyond those that satisfy (74).  
In this case the
diffeomorphism constraints have
the standard local form,
\f
0= {\cal D}(v)= \int_\Sigma v^a F_{ab}^i\tilde{\Pi}^{bi} .
\ff
Similarly, we can define the Hamiltonian constraint.  Due to the 
constraint
on the trace of $\Phi_{ij}$, it is easy to see that the following 
is, for all
smooth $N(x)$ on $\Sigma$, an identity,
\f
0= \int_\Sigma N\left (
\epsilon_{ijk}\epsilon_{abc}\tilde{B}^{ai}\tilde{E}^{bj}\tilde{E}^{ck}
+ \lambda det( \tilde{E})
\right )
\ff
where $\tilde{E}^{ai}\equiv \tilde{B}^{bj}(\Phi^{-1})_{ij}$.  We 
then may
attempt to write a constraint for all $N$, by using the 
definition of the
momenta (53) and so writing, 
$\tilde{E}^{ai} = \tilde{\Pi}^{ai}-(3/\lambda ) \tilde{r}^{ai}$, 
where the
latter stands for the surface term,
\f
\tilde{r}^{ai}= \int_{\partial \Sigma}d^2S^{ab}(\sigma )
A_b^i (\sigma ) \delta^3 (x, S(\sigma))
\ff
However, one quickly sees that for all $N(x)$ the 
resulting functional
is singular, due to the quadratic and higher terms 
in $\tilde{r}^{ai}$.
Thus, we may only take as  generators of gauge transformations
those constraints with $N(x)$ vanishing on the boundary.
Thus, our Hamiltonian constraint is simply the standard local one,
\f
0= {\cal C}(N) = \int_\Sigma N\left (
\epsilon_{ijk}\epsilon_{abc}\tilde{B}^{ai}
\tilde{\Pi}^{bj}\tilde{\Pi}^{ck}
+ \lambda det( \tilde{\Pi})
\right )
\ff
with the condition that 
\f
N|_{\partial \Sigma } =0
\ff

\subsection*{Vd.  The hamiltonian}

We may ask also if we can extend this constraint to a Hamiltonian,
which will give us a nontrivial time evolution, as measured by a 
lapse
function that reaches to the boundary.  We may do this in the 
standard
way, by noting that we may add a boundary term to the 
Hamiltonian
constraint, so that the resulting functional is now functionally 
differentiable
for all $N$ including those that are nonvanishing at the boundary.  
This,
not surprisingly, takes the same form as in the asymptotically flat 
case,
\begin{eqnarray}
{\cal H}(N)&\equiv &
\int_\Sigma N\left (
\epsilon_{ijk}\epsilon_{abc}\tilde{B}^{ai}
\tilde{\Pi}^{bj}\tilde{\Pi}^{ck}
+ \lambda det( \tilde{\Pi})
\right )
\nonumber \\
&&-\int_{\partial \Sigma}d^2S_a N A_b^k 
\tilde{\Pi}^{ai}\tilde{\Pi}^{bj}
\epsilon_{ijk}
\end{eqnarray}
This boundary has the property that when taken to infinity in the
presence of asymptotically flat boundary conditions, it goes to the
ADM mass.  However, in the presence of these boundary conditions
there is no such simplification, so this remains a non-linear term.

\subsection*{Ve.  Some observables and symmetries 
associated with the boundary}

As I indicated in the introduction, one motivation for 
introducing a finite boundary in quantum gravity 
is to have an infinite dimensional
algebra of observables, associated with measurements that
might be made on the boundary of the system.  We have
already in Chapter 3 mentioned the algebra of observables
${\cal A}^{boundary}_{\cal S}$.  Here we introduce them
in the context of the classical canonical theory we have
just defined.  The first are the loop algebra $T[\alpha ]$, the 
traces of holonomies of loops $\alpha$
in the boundary.    We may note now that,
under the conditions just
derived, these commute with
all the constraints of the theory, and so are a subalgebra
of the algebra of physical observables of the theory.

The algebra
of these observables is defined by the Dirac brackets (66).
Further, because we are working
with the Euclidean boundary conditions this algebra is
a star algebra, because
\f
T[\alpha ]^* = T[\alpha ]
\ff

The second set of observables is the areas of regions of the
boundary, which we denote by 
${\cal A}[{\cal R}] =\int_{\cal R} \sqrt{h}$  where
$h_{\alpha \beta}$ is the two metric induced on
the boundary $\cal S$.  These are also real.

Finally, the third set of observables is given by the generators of 
the diffeomorphism group of the boundary 
$Diff(\partial \Sigma )$,
which as we have emphasized is a symmetry group. The
generators of $Diff(\partial \Sigma )$,  may be denoted,
by $d(v)$, for $v^\alpha$ a vector field on the boundary.  
They are given, 
as in the full theory, by
\f
d(v) \equiv 
\int_{\partial \Sigma}  v^\alpha (F_{\alpha \beta}^i 
\tilde{\Pi}^{\beta i} + F_{\alpha r}^i \tilde{\Pi}^{ri})
\ff
Again, by the reality conditions,
\f
d(v)^*=d(v)
\ff

\section*{VI.  Quantization in the presence of boundaries}

We are finally in a position to construct the quantum
theory that corresponds to the classical theory we have
constructed in the last two sections, and recover the results
we described in the section III.

I give here an outline of the main steps in the quantization,
stressing those aspects that are new in this case.
Full details will appear elsewhere.  

We proceed in the following order.  First, I will review a few more of
the basics of  quantum Chern-Simons theory on
the two manifold with punctures, 
$({\cal S},y_\alpha , j_\alpha )$.  Then we will extend this
first to the kinematical and then to the diffeomorphism
invariant quantum theories on the interior of $\Sigma$.

\subsection*{VIa.  More facts about quantum Chern-Simons theory}

Before proceding to the quantum theory in the whole
manifold $\Sigma$, we need to fill in a few of the details
of quantum Chern-Simons theory on 
$({\cal S},y_\alpha , j_\alpha )$ that were left out
of the sketch in section 2.  The main fact we will 
need is that

{\bf Spin network basis:}  An orthonormal basis 
for the state space 
${\cal H}^{CS}_{{\cal S}, y_\alpha , j_\alpha }$ is given
by the set of independent $q$-spin network trees
in $\cal S$ in which one edge emerges from each
puncture $y_\alpha$ colored with spin $j_\alpha$.
By independent here we mean under the equivalence
relations of quantum spin networks and homotopy
on ${\cal S}-y_\alpha $.  The key equivalence relation
among the spin networks are the defining relations
of quantum $6-j$ symbols, which tell us that the four
valent node can be decomposed in two equivalent ways
as a product of two trivalent nodes, analogous to the usual
duality diagrams in which annihilation and creation processes
are equivalent to scattering.

For finite $k$ and genus $n$, these spaces are finite
dimensional.

This is a standard result of conformal field 
theory\cite{moreseiberg,louis2d3d}.
It is sometimes stated in a language or trinions, in
which the manifold ${\cal S}-\{ y_\alpha \}$ is decomposed
according to trinions, colored with representations of
$SL(2)_q$ on the three boundaries, so that addition
of $q$-angular momentum is 
satisfied\cite{moreseiberg,louis2d3d}.  The description
in terms of trinions is completely equivalent to that of
quantum spin networks.  

To see this we may recall that a trinion
is a sphere with three disks removed, which are labeled
by representations.  A trinion decomposition of a punctured
surface without boundary may be given in which the three
boundaries of each trinion are associated with either
punctures or circles along which the surface has been cut.
In the latter case, the two labels on the circles of each
two trinions are joined must match.

Given a trinion decomposition of the surface, we may associate
a spin network by drawing a graph on each trinion which
consists of three lines coming from each boundary joined
at one trivalent vertex.  At any circle where two trinions
are joined we join the two lines going out to those circle,
which we may since the representations match.  Correspondingly,
it is straightforward to see that given any spin network
on the surface we may, under the equivalence relations available
for flat connections, be replaced by a sum over trivalent
spin networks, so that each may be replaced by an equivalent
trinion decomposition.

Given this picture of the 
state spaces, the action of the elements of the observable algebra
${\cal A}_{{\cal S},y_\alpha , j_\alpha}$ may be computed
following the standard methods of conformal field theory.

This algebra is generated by the generators of
$Diff({\cal S}-y_\alpha )$ and by $t[\alpha ]$,
where $\alpha \in \pi^1[{\cal S}-y_\alpha ]$.  The algebra
is given by (7), together with the action of the diffeomorphisms
on homotopy classes.

\subsection*{VIb.  Kinematical quantum theory}

We now begin the construction of the kinematical
quantum theory on the whole manifold $\Sigma$.
We begin in the connection representation,  later,
when we have established the kinematical quantization
we will transform to the loop representation to find
the diffeomorphism invariant states, following the
usual ideas for manifolds without 
boundary\cite{lp1,carlo-review,ls-review,spinnet-us}.

We will seek to follow the  constructions
that underlie the application of the loop representation
to the non-perturbative quantization of diffeomorphism
invariant theories in the absence of boundaries.  We will,
however, have to modify the construction at several
points to account for the conditions on the boundaries.

\subsection*{VIc.   The spin network 
basis in the connection representation}

The basic idea of the connection representation is that
the states are functions of $A_{ai}$ and the canonical
momenta $\tilde{\Pi}^{ai}(x)$ are represented
by $\delta /\delta A_a^i (x)$.
As usual, the crucial element in the
quantization is the treatement of the
Gauss's law constraint. There are two main approaches
to quantization with constraints, quantize and then apply
the constraints as operator equations, or solve the
constraints classically and then quantize.
Most of the work that follows is performed using the
first method.  But, as a check, in subsection {\bf VIf } 
the same
results are derived using the solve and 
then quantize approach.

We choose to first quantize, and then solve this on the
quantum states.
We may note that
acting on states in the connection representation the Gauss's
law constraint gives, 
for every $\Lambda^i$ on $\Sigma$,
\begin{eqnarray}
\hat{\cal G}(\Lambda) \Psi [A] &=& -  
\int_\Sigma d^3 x {\cal D}_a \Lambda  (x) )^i 
{ \delta \Psi [A] \over \delta A_a^i (x) }
\\
 &&-{1\over 2} 
\int_{\partial \Sigma} d^2S^{ab} (\sigma ) \Lambda^i (S(\sigma ))
\left (  \epsilon_{abc} { \delta \over \delta A_c^i (x) }+
{3 \over \lambda } F_{ab}^i (S(\sigma )) 
\right ) 
\Psi [A] =0
\nonumber
\end{eqnarray}
 
It is important to note that because this must be true for
every choice of $\Lambda^i$, the volume and surface terms
must be solved separately.  This is because we must solve
it for every $\Lambda^i$ with support that excludes the 
boundary as well as for every choice of support  on a region
that extends an arbitrarily small distance $\epsilon$ into
$\Sigma$ from the boundary.

We may then proceed to solve the volume term in the
Gauss's law as usual by expressing the states in terms of
loops, corresponding to Wilson loops of 
the connection $A_a^i$  on $\Sigma$.  We then assume
that given a set of loops $\gamma_I$ we may
construct an overcomplete basis of 
states in the connection representation
of the form
\f
\Psi_\gamma [A]= <A|\gamma_I > = 
\prod_I T[\gamma_ ,A]
\ff
where the $T[\gamma_I , A ]$ is the trace of the
holonomy in the fundamental representation.
The loops $\gamma_I$ may enter or leave
the boundary $\cal S$, and run along in it. 

These loop states are overcomplete because of the
Mandelstam and retracing 
identities\cite{aa-review}.  However
an independent basis may be 
constructed\cite{spinnet-us,otherspins}
which is in one to one correspondence with the
spin networks.  This is, of course, a standard
result in lattice gauge 
theory\cite{lgt} and topological
field theory\cite{louisdavid,louisigor,ooguri}, we review the basic 
idea here.

Each such basis element
is equal to a linear combination of loop states which is constructed
in the following way (for details, see \cite{spinnet-us}).    
A spin network 
diagram is a spin network, as defined in the previous section,
together with an orientation which is given by an imbedding
of the network in the plane.  
To each edge of the network, labled by an
integer $p$, associate $n$ copies of that segment.  To each vertex
of valence higher than three, there is also a label, which indicates
how the routings through it are to be decomposed in terms of
trivalent vertices (this is also given uniquely in terms of the
imbedding in the diagram).  Then combine all these
segments joining them across the vertices in all possible
ways, resulting in a finite set of multiloops $\gamma_i$, 
in which each
edge is traced the number of times given by its labeling.
The spin network basis is then a sum over the basis elements
labled by these  multiloops.  In the connection representation
it is given by
\f
\Psi_\Gamma [A] = \sum_{\gamma_i \in \Gamma} (-1)^{r_i} 
T[\gamma_i ,A]
\ff
where the $T[\gamma_i ,A] $ are the usual Wilson loops in
the fundamental representation and the 
overall sign $r_i$ is defined by a rule which depends on
the imbedding of the graph in the plane.

As explained in \cite{spinnet-us}, once a 
unique rule for the labeling of the
higher valence vertices is given, the elements of the spin network
basis are independent, in that they satisfy no further identities.  
 
\subsection*{ VId.  Restrictions due to the Dirac brackets 
at the boundary}

There is, however, an important restriction on the states of
the form (86) or (87), when the 
edges of the spin network run in the 
boundary $\cal S$.  This is that, because of the commutation
relations (66), the connections restricted to the boundary do not
all commute with each other.  If we allow states to have arbitrary
dependence on the connection on the boundary we will have
too many states, as we should have enough to give a basis
of functions on the configuration space, and not on the
whole phase space.

Several different ways to accomplish this have been developed
in studies of Chern-Simons theory.  One which
has been often used 
depends on the trinion decomposition we discussed
in the previous section.   On each trinion we may pick a 
complex coordinates and require that the states 
are holomorpic functions of $z$.  Holomorphic
functions may be constructed on the whole manifold by insuring
that the holomorphic conditions can be continued accross any
boundary joining two trinions.

For our purposes we may make the following equivalent
construction.
We may make a choice
of local coordinates $r, \theta$  in a neighborhood of each
puncture such that curves of constant $r$ circle the puncture,
while curves of constant $\theta$ run radially away from it.
On each trinion, these three patches may be joined consistently
to yield coordinates on the whole trinion.  Furthermore,
the coordinates can be matched along each boundary
where two trinions meet, as these are circles of constant $r$.
We may then
require that the states are functions only of the $A_r^i$
components of the connections.  This means that the
curves in the spin network states (87) can run on the trinions
only along the radial directions, on constant $\theta$ curves.
This allows us to have a trivalent spin network on each
trinion, where the three edges run from each disk radially
to a point on the trinion where the three coordinate patches
meet.  We will call a graph $\Gamma$ on $\cal S-\{ y_\alpha \}$, 
with ends on the
punctures, an allowable graph, when there
is a trinion decomposition of the surface, and a choice of
local coordinates on each trinion, such that the edges of the
graph that run in the surface only run in the radial coordinate
direction in those coordinates on each trinion.

We will assume that the spin network states then depend
only on allowable graphs, when they run in the surface.

\subsection*{VIe.  
Imposing the surface term in the Guass law constraints}

The next step is to impose the boundary term in
the Gauss law constraint, eq. (85) on the space of
states we have just described.  To do this it will be
useful to decompose the kinematical state space
in terms of sets of points and representations,
$(y_\alpha , j_\alpha )$ on $\cal S$, as we described
in section III.   We then may study the action of
the surface term in (85) on states in the
subspace ${\cal H}^{kin}_{y_\alpha , j_\alpha }$.
This consists of all spin network states of the
form (87) where the network $\Gamma$ enters
the boundary only at the marked points $y_\alpha$,
and so that the edges of $\Gamma$ that intersects
$\cal S$ at $y_\alpha$ carries the representation
$j_\alpha$.

We will now see that in each such sector, the action
of the surface part of the Gauss's law constraint is to
to impose exactly the Gauss's law constraint in the
Chern-Simon's theory with sources at the points
$y_\alpha$ with spins $j_\alpha$.

We then consider, for all functions
$\Lambda^i$ with support that includes 
a piece of the boundary $\cal S$, the condition,
\f
\int_{\partial \Sigma} d^2S^{ab} (\sigma ) \Lambda^i (S(\sigma ))
\left (  \epsilon_{abc} { \delta \over \delta A_c^i (x) }+
{3 \over \lambda } F_{ab}^i (S(\sigma )) 
\right ) \Psi_\Gamma  [A] =0
\ff
where the state can only have support on allowable graphs 
$\Gamma$.
We may consider first of all the case
that the support of the support of  $\Lambda^i$ does not
include any of the punctures $y_\alpha$.  By considering
all such cases, we reach the conclusion that, in the
connection representation, the states must have support
only on flat connections in ${\cal S}-y_\alpha^\Gamma $.
This is because the $\delta / \delta A_{ai}$ terms annihilate
such states.  The curvature $F_{ab}^i$ can be
represented in terms of the holonomy of a loop that encloses
no punctures, the constraint then says that the holonomy must
be trivial, which means that the connection is flat.

This means that the loop
function corresponding to $\Gamma$  
can only be functions of the homotopy of those 
portions of the loops
that run in ${\cal S}-y_\alpha^\Gamma $.  (Both of these
results are found using the standard arguments developed
for $2+1$ gravity and Chern-Simons 
theory\cite{witten-tqft,cs-other,ls-johnshopkins}).
This means that the states can only depend on the
spin networks up to homotopy in the boundary and the
equivalence relations among spin networks.  

We have then reached the following result:  Fix a spin network 
$\Gamma^\prime $ with open edges, 
in the interior of $\Sigma$ such that the open ends
touch the boundary at a set of
points $y_\alpha$, with edges labeled by representations
$j_\alpha$.    For every independent way to complete 
$\Gamma^\prime$ to a spin network 
$\Gamma$
with the ends joined
by edges running in the boundary, we will have an independent
kinematical quantum state of the gravitational field in
all of $\Sigma$.  What we have just shown is that these
independent possibilities are exactly given by the ways to
draw a trivalent spin network in the surface $\cal S$ with
ends at $y_\alpha$ labeled by the $j_\alpha$, up to homotopy
of the loops in ${\cal S}-\{ y_\alpha \}$ and the $6-j$
symbol relations among trivalent spin networks.   
What we have done is reproduced the 
description of the state space of Chern-Simon theory 
associated to $({\cal S},y_\alpha ,j_\alpha )$, with one 
difference, which is that we have not so far obtained the
restriction to quantum spin networks at level $k$.  

To realize this we must take the last step and impose
the remaining constraints, which are the surface
parts of the Gauss's law in the case that $\Lambda^i$
has support on a region of the boundary that includes
at least one puncture $y_\alpha$.  It is, of course,
sufficient to restrict $\Lambda^i$ to have support
on a region that includes a single puncture $y_1$.

We should proceed with care, as 
$F_{\alpha \beta}$ is now a non-trivial operator
acting on the states.  
$F_{\alpha \beta}$ involves products of operators that
don't commute with each other, and a complete discussion
of the relation (88) must involve a careful regularization procedure.
To establish the necessity of the restriction to quantum
spin networks, however, it will be sufficient to compute
the simplest case, which is the case the puncture is in
the fundamental representation, so $j_1=1/2$, in the 
limit of large $k$.

To proceed, we consider how we may define the
action of a loop operator $\hat{T}[\alpha ]$, for a loop
$\alpha$ that goes once around the puncture at $y_1$
at constant $r$ in some choice of radial coordinates.
We may note that this may be defined two ways, by realizing
directly the Poisson bracket relations (66) on the states 
or by using the surface constraint (88) for a region $\cal R$
which is the interior of $\alpha$.  The basic point,
as we will see, is that
the surface constraint must define the action of a loop
operator that surrounds one puncture.

Note that as all the angular components, $A_\theta^i$  commute
with each other there is no ordering problem in defining
$\hat{T}[\alpha ]$.  But there are ordering issues when evaluating
its effect on states based on admissable loops.

We begin with the second method.  To construct an operator
to represent $T[\alpha ]$, we may use the non-Abelian
Stokes theorem \cite{nas} to write
\f
\hat{T}[\alpha ]\Psi_\Gamma [A] =
Tre^{ \imath \int_{R} d^2S^{ab} (\sigma ) U_{\eta_\sigma} 
\hat{F}_{ab}(\sigma) U_{\eta_\sigma^{-1} }  } 
\Psi_\Gamma [A]
\ff
Here $\eta_\sigma$ is a loop from an arbitrary base point
somewhere on $\alpha$ to the point $\sigma \in {\cal R}$.
The actual loop generally is defined in a particular way through
the construction of the Non-Abelian Stokes theorem, but using
the fact that the connection is flat on ${\cal R} -y_1 $ we
may take any loop that connects them with trivial homotopy.

We may now use the surface constraint from Gauss's law (88).
Restricting to the case that there is a single $j=1/2$ line
meeting the manifold at $y_1$ in $\cal R$, we have
$\Psi_\Gamma [A]=T[\gamma ,A] \xi[A]$, where $\xi$ is not
acted on by the loop operator $T[\alpha ]$.  We then want
to evaluate,
\f
\hat{T}[\alpha ]T[\gamma, A] =
Tre^{-{2 \pi \over k} \int_{R} d^2S_c (\sigma ) U_{\eta_\sigma} 
\tau^i U_{\eta_\sigma^{-1} }{ \delta \over \delta A_c^i (\sigma )  }}
T[\gamma , A]   .
\ff

It is straightforward to evaluate the first two terms in
the expansion of the exponential.  The term in $1/k$ vanishes
because the trace of a symmetrization gives zero, so that the
leading non-trivial term is order $1/k^2$. After
a simple calculation, using the methods described 
in \cite{spinnet-us},  
we find that
\f
\hat{T}[\alpha ]T[\gamma, A] = \left [ 2- {3 \over 2} 
\left ( {2\pi \over k}  \right )^2 + O(k^{-3}) \right ] T[\gamma , A]
\ff
There is a problem with the normalization of this result, as we shall
see in a moment.  The basic problem is that it gives a non-zero
answer for the action of the operator, even in the case in which
the loop $\alpha$ surrounds no punctures.  However, in order to 
realize
the commutation relations (66) in a way that is consistent with the
restriction that the states in (87) are admissable, according to the
rules we developed above, in this case the action of the operator
must vanish.  To see this, let us consider the case in which
we want to realize the commutation relations coming from (66):
\f
[ \hat{T}[\alpha ], \hat{T}[\gamma ] ] = {2\pi \imath \over k}
Int[\tilde{\alpha},\tilde{\gamma} ] \left (
\hat{T}[\alpha \circ \gamma ] - \hat{T}[\alpha^{-1} \circ \gamma ]
\right )
\ff
in the case that, as here, $\alpha$ is not an admissable loops,
but $\gamma$ is.  We may note that as $\alpha$ is not
admissable, the corresponding operator
$\hat{T}[\alpha ]$  cannot create a loop, as in the usual definition
of the Wilson loop operator.    We can define
this action according to the commutation relations (66), in
the following way: a loop operator is treated as a $T^0$
when the loop is admissable, and therefor adds that loop,
while it acts like the  standard definition of a $T^1$ in the
case that the loop is non-admissable (and in the surface
$\cal S$). ,  This will guarantee that the
algebra (66) is realized win all cases.
We then have, in the case that only $\gamma$ is admissable,
\f
\hat{T}[\alpha ] \Psi_\gamma [A] = 
{2\pi \imath \over k}
Int[\tilde{\alpha},\tilde{\gamma} ] \left (
\Psi_{\alpha \circ \gamma}[A] - 
\Psi_{\alpha^{-1} \circ \gamma}[A]
\right )  .
\ff
Clearly, we can only require that this is true if it is the case that
all three of the loops $\gamma$, $\alpha \circ \gamma $ and
$\alpha^{-1} \circ \gamma $ are homotopic to admisable
loops in the region ${\cal R}-y_1$.  We shall see in a moment
that this is the case, if $\gamma$ is admissable.

There is, however, a subtle point, which is that the definition
we have just given requires that $\hat{T}[\alpha ]$ annihilate
a loop state $\Psi_\gamma [A]$ in the case that $\alpha$ is
not admissable and does not intersect with 
and $\gamma$ in the surface.  This disagrees
with the result of the non-Abelian Stokes theorem, which
is (91). To make them
consistent with each other,  we must take into account that
the classical non-Abelian Stokes theorem may have to be 
modified when it is applied to quantum operators that don't
commute with each other.  To account for the discrepency,
it must then be that we missed a constant subtraction in the 
evaluation of the non-Abelian Stokes theorem coming from the
fact that it involves manipulation of non-comuting operators.
To make it match at leading order, we must then modify 
(90) so that
\f
\hat{T}^\prime [\alpha ]T[\gamma, A] =\left [
Tre^{-{2 \pi \over k} \int_{R} d^2S_c (\sigma ) U_{\eta_\sigma} 
\tau^i U_{\eta_\sigma^{-1} }{ \delta \over \delta A_c^i (\sigma )  }}
-2 \right ]
T[\gamma , A]
\ff
where the $\prime$ indicates that a subraction has been
done in the definition of the operator so that it agrees with
the requirement that when $\alpha$ is a non-admissable
loop $\hat{T}[\alpha ]$ must annihilates 
states supported on loops that
have no intersection with $\alpha$ in the boundary.  We then have,
\f
\hat{T}^\prime[\alpha ]T[\gamma, A] = \left [ {3 \over 2} 
\left ( {2\pi \over k}  \right )^2 + O(k^{-3}) \right ] T[\gamma , A]
\ff

We may now compare this answer to the one we get if
we directly use the definition  (93) of the action of the
$\hat{T}[\alpha ]$ operator to 
evaluate the action of the loop operator on the spin
network states.   

Now we come to a second subtelty, which is indeed the key point 
of the whole calculation.  This is whether or not we allow
homotopic equivalence to include smooth deformations of the
loops in ${\cal S}- \{ y_\alpha \}$  that rotate one or more
of the punctures by $2\pi$. 
Note that if we take the naive picture in which
the puncture may be twisted, the loops $\gamma\circ \alpha $
and $\gamma \circ \alpha^{-1}$ are homotopic to each other
and to $\gamma$ in ${\cal R}-y_1$. Hence, under this assumption,
both the first and the second terms of (93) are equal to each other,
so that the action
is zero.  This means that the assumption that homotopy classes
include those in which the puncture is rotated is 
inconsistent with the surface constraints (88), because 
we have
just performed the calculation another way that uses them 
and found a non-vanishing answer.

This means that the loops must be framed, so that there is a factor
coming from twisting the puncture.  We must then have,
\begin{eqnarray}
\Psi_{\gamma\circ \alpha}[A]  &=& e^{{2\pi \imath \theta \over k}}
\Psi_{\gamma }[A]
\nonumber \\
\Psi_{\gamma\circ \alpha^{-1}}[A]  &=& e^{-{2\pi \imath \theta 
\over k}}
\Psi_{\gamma }[A]
\end{eqnarray}
where the factor $\theta$ must come from the operator ordering
that requires the loops to be framed.  We then have,
\f
\hat{T}^\prime [\alpha ]T[\gamma, A]= {2\pi \imath \over k} 
(2\imath )
sin({2\pi \theta \over k})
\ff
Comparison of this at large $k$ with (95) tells us that
we must have $\theta =-3/4$.  We may note that this factor
may be computed directly in Chern-Simons 
theory\cite{witten-tqft,ls-cs} and one
gets precisely this same value of $\theta$.    Indeed, the 
action of a twist is one of the defining relations of the Kauffman
bracket, so that this factor is determined directly by the
Jones polynomial\cite{qnet}.  It is non-trivial that this 
calculation has
led to the same value of $\theta$.

We have thus shown that to leading non-trivial order in
$k$, the loops must be framed in agreement with the result
in Chern-Simons theory.  Furthermore, we have checked directly,
by taking into account the surface constraint (88) that to
leading non-trivial order in $k$, the operator associated with
a loop that surrounds a puncture agrees precisely with the
value in Chern-Simons theory, at least in the case
that the puncture has $j=1/2$.  Thus, for this
same case, we have deduced the
isomorphism (18) of the observables algebra of Chern-Simon
theory on ${\cal S}- \{ y_\alpha \}$ with the subspace of
the surface algebra of quantum gravity
${\cal A}^{\prime gr}_{{\cal S}, y_\alpha , j_\alpha }$

This derivation will be carried out in the general case elsewhere.
For now, we may check that the same result
follows by taking the other approach to quantizing in which
we first solve the Gauss's law constraint, and then quantize.

\subsection*{VIf.  Constraining and then quantizing}

We have run into a cumbersome problem, that involved
having to define an operator by doing a subtraction,
because we
have followed the route of imposing the Gauss law constraints on
the states.  While we have been able to extract the result
by considering the leading terms at large $k$, 
there is an easier way to reach the same conclusion.
In most treatments of Chern-Simons theory, the
problem of defining the Gauss law constraint as an operator
equation in a way that is consistent with the commutation
relations (66) is avoided by first solving the
constraints classically, which yields the reduced phase
space, and then quantizing.  When this path is followed
in the case of punctures, the problem is how to describe
the effect of the puncture in the classical phase space.
This is usually done by adding a degree of freedom to 
each puncture, so that the Gauss's law constraint becomes
\f
{k \over 4\pi } \epsilon^{\alpha \beta }F_{\alpha \beta}(\sigma )
 = \sum_\alpha r_\alpha  \delta^2(\sigma, y_\alpha )
\ff
where the $r_\alpha $ are classical variables, valued
in the Lie algebra of $SU(2)$ associated to the 
punctures at $y_\alpha$.
In order that the classical variable $r_\alpha$
will yield a state in
the $j_\alpha$'th representation on quantizing, it is taken
to lie in the coadjoint orbit, with a symptectic structure given
by a choice of representation, using the connection between
symplectic structures on groups and their 
representations\cite{witten-tqft,cs-other}.
For our purposes, this means that, up to a gauge transformation,
$r_\alpha$ must be the equal, in the Lie algegra, 
to the root that is
the highest weight of the $j$'th representation.   
This means that we
may take 
$r_\alpha = \omega_\alpha \lambda_\alpha  \omega^{-1}_\alpha$, 
where
$\omega_\alpha \in SU(2)$ and $\lambda_\alpha$ is a root
associated with the spin $j_\alpha$ representation.

Given the classical system described by (98), 
quantization then yields the
Chern-Simon theory as we described it 
above\cite{witten-tqft,cs-other}.

We may follow this development by finding a suitable classical
problem that may correspond to our constraint (88).  The main
question here, as in the case of Chern-Simon theory, is what the
representation of the puncture will be in the classical theory.

There is readily available an answer to this, which is to recall
that, as described in \cite{ls-review,discrete}, the classical 
limit of the 
kinematical states in the loop
representation correspond to a certain kind of discrete 
geometry, which is analogous to the Regge calculus.  However,
in this case, the metric geometry is distributional, with
the frame fields having support only on the edges of the graphs.
We may note that this is appropriate, because the classical limit
must only yield something that approximates a smooth geometry
in the case that the state is based on a large graph such as the
weaves described in \cite{weave}.  We will see that if we use this
classical picture, we reproduce the constraint (98) of Chern-Simons
theory in the case of punctures.

We may then consider the classical geometry that corresponds to
a spin network $\Gamma$ to be given by a distributional frame
$\tilde{E}^{ai}_\Gamma$ defined by,
\f
\tilde{E}^{ai}_\Gamma (x) = G \sum_l 
 \int ds \delta^3 (x, \gamma_l (s))
\dot{\gamma}_l^a (s)  t^i_l (s)
\ff
Here, $l$ labels the edges of the graph and $t^i_l(s) $ are elements
of the Lie algebra.  The magnitude of these Lie algebra elements
must be picked by requiring that the geometric observables
computed using this classical frame field agree with the  
eigenvalues of the corresponding quantum operators  in the
spin network state $\Gamma$.  It is interesting that the
area and volume observables may be computed for such
frame fields, in spite of their distributional character, using
exactly the same regularization technique as is employed in the
construction of the action of the quantum operators on the 
loop states\cite{ls-review}.  The result is that the classical 
correspondence
principle 
applied to the area observables 
requires that $|t_\alpha | = \sqrt{j_\alpha (j_\alpha +1)}$.
This means that the $t_\alpha$ must be taken so that
\f
t_\alpha (s) = \omega (s) \lambda_\alpha \omega^{-1} (s).
\ff
where $\omega (s)$ is an $SU(2)$ valued field on the edge.
If we now plug this into the surface 
term of the constraint (68) we find
exactly the condition (98) with $t_\alpha =r_\alpha$.

Thus, from considerations of the classical limit of the
quantum gravity theory, we arrive at the same classical
version of the surface Gauss's law constraint which is used as
the starting point of the quantization of the corresponding
Chern-Simons theory.  This tells us that the isomorphisms of
the observable algebras (18) must hold at the classical level.

Quantization must then yield an algebra of quantum operators
at the surface which is identical to the algebra of observables
of the Chern-Simons theory.  Thus, the two representation
spaces must be identical, which means that 
the restriction to the quantum spin networks must apply
also in the quantum gravity case, with the identification of the level
$k$ coming from quantum gravity given by (49) corresponding to
the $k$ of the classical Chern-Simons theory.  
Further more, as the level is renormalized in the Chern-Simons
theory, but the (49) relation holds between the classical
theories, this means that $k$ will be renormalized from 
$k$ to $k+2$ also in the quantum gravity case.

\subsection*{VIg.  Kinematical inner product}

We may now use what we have learned to construct a kinematical
inner product for the quantum theory, that respects the
reality conditions of the Euclidean theory.  It is sufficient
to impose an inner product on each subspace 
${\cal H}^{kin}_{y_\alpha ,j_\alpha}$ because we
know that states in different of these subspaces must be
orthogonal as they are each eigenspaces of the operators
that measure the areas of surfaces.
The results of the last subsection mean that we may
decompose each of these spaces according to the analogue
of (24)
where we now have a characterization of the kernal:
${\cal K}^{kin}_{{\cal C},y_\alpha ,j_\alpha}$ must
be spanned by states corresponding to each 
quantum spin network
$\Gamma$ in the interior of $\Sigma$ with open ends
with labels $j_\alpha$ that meet the boundary at the
points $y_\alpha$.  

We may now impose a kinematical
inner product $<|>_{{\cal C},y_\alpha ,j_\alpha}^{kin}$
in each of these subspaces.  This must satisfy a
requirement that every area and
volume operator, associated with every surface and
region in the interior of $\Sigma$ orthogonal. But given
any two distinct spin networks $\Gamma$ and $\Gamma^\prime$
in the interior there will exist such operators such that the
states $\Psi_{\Gamma}$ and $\Psi_{\Gamma^\prime}$ are both
eigenstates, but with distinct eigenvectors.  This means
that all states associated with distinct networks must
be orthogonal to each other, so that the inner product
must have the form,

\f
<\Gamma |\Gamma^\prime>^{kin}_{{\cal C},y_\alpha ,j_\alpha}=  
n_\Gamma 
\delta_{\Gamma \Gamma^\prime}
\ff
The restriction to this class at the kinematical level is
justified by the fact that the spin network
states are eigenstates of a complete set of
commuting observables, which are the volumes of
arbitrary regions  and the areas of
arbitrary surfaces in $\Sigma$ \cite{volume}.  Further the distinct
spin networks are completely distinguished by
their respective eigenvalues.  Therefor they must
be orthogonal to each other in any inner product 
chosen to respect the hermiticity of these observables.

We may note that additional observables must
be invoked to fix the coefficients $n_\Gamma$.
In the case of Euclidean signiture, with which
we are now concerned, it is a simple exercise
to show that imposing that the loop
operators $\hat{T}^0[\gamma ]$ be hermitian
leads to the conclusion that all the $n_\Gamma$'s
are equal to unity\cite{ham2,gangof5}.

To complete the description of the kinematical state
space, we may write an orthonormal basis,
\f
|\Gamma , z >= |\Gamma > \otimes |z >
\ff
where $|z>$ 
are a basis in ${\cal H}^{CS}_{{\cal C},y_\alpha ,j_\alpha}$.
But we know that a basis for the states of the Chern-Simon theory
are given by the independent trivalent quantum 
spin networks in the surface with ends
on the punctures.  Thus, we have the following construction.
Let $\Gamma$ be a quantum spin network in $\Sigma$ which
meets the boundary $\cal S$ at points $y_\alpha^\Gamma$
with quantum spins $j_\alpha$ and is admissable and
trivalent when it runs
in the boundary.  Then it induces a state 
$|z^\Gamma > \in 
{\cal H}^{CS}_{{\cal C},y_\alpha^\Gamma ,j_\alpha^\Gamma}$
given by the quantum spin network that is 
its intersection with the boundary.  It also induces a quantum
spin network $\tilde{\Gamma}$ that runs in the interior
of $\Sigma$ and meets the boundary at the points 
$y_\alpha^\Gamma$.  Then the kinematical inner product of
the Euclidean theory is
\f
<\Gamma_1 |\Gamma _2 > = < z^{\Gamma_1}|z^{\Gamma_2} >_{CS}
 \delta_{\tilde{\Gamma}_1 \tilde{\Gamma}_2 }
\ff

 We may note that this inner product is nonseperable.  This would
no doubt be a problem, were we doing anything other than
quantizing a diffeomorphism invariant theory.  We will see that
this problem is fixed at the next step.

\subsection*{VIh.  Diffeomorphism invariant quantum theory}

Finally, we may next construct the diffeomophism invariant
states.  As the diffeomorphisms are frozen on the boundary, this
is straightforward, spin networks $\tilde{\Gamma}$ that meet
the boundary are replaced by their diffeomorphism invariant
classes following the logic of \cite{lp1}.  The fact that the 
diffeomorphisms are fixed on the boundary means that 
the subspaces
${\cal H}^{diffeo}_{y_\alpha ,j_\alpha }$ and 
${\cal H}^{diffeo}_{\phi \circ y_\alpha ,j_\alpha }$, where
$\phi$ is a diffeomorphism of the boundary are physically
distinct, and are related by a symmetry transformation.   We then
have diffeomorphism invariant states $|\{ \Gamma \} >$,
which correspond to an admissable, trivalent loop, running
in ${\cal S}-\{ y^\Gamma_\alpha \}$, together with a diffeomphism
equivalence class of continuations $\{ \tilde{\Gamma} \}$
into the boundary, where, again, the diffeomorphisms vanish
on the boundary.  The diffeomorphism invariant inner product
is then
\f
< \{ \Gamma_1 \} |\{ \Gamma_2 \} > = 
\delta_{y^{\Gamma_1}_\alpha y^{\Gamma_2}_\beta }
< z^{\Gamma_1}|z^{\Gamma_2} >_{CS}
 \delta_{\tilde{\{ \Gamma}_1 \} \{ \tilde{\Gamma}_2 \}}
\ff

\section*{VII.  Conclusions}

In closing, it may be interesting to mention several directions
that may be developed beginning with the results described
here.

A crucial question is whether it is possible to extend these
results from the Euclidean to the Minkowskian case.  As was
mentioned above, there is not necessarily a problem with
applying the self-dual boundary conditions in the case of
Minkowskian signature, as only one third of the self-dual
conditions are imposed on the boundary.  What Minkowskian
solutions are consistent with these conditions needs to
be investigated.  The possibility that these results might
also apply to the Barbero formalism\cite{fernando-real} should
also be investigated.

An especially interesting suggestion is whether these or related
conditions can be achieved on an apparent horizon.  If so,
the results of Carlip\cite{carlip}, who fixes boundary 
conditions on
horizons in $2+1$ gravity, may be brought to the physical
$3+1$ dimensional case.

The main issue to be faced in extending these results to the
Minkowskian case is that the results I used from quantum Chern-Simons
theory depend to some extent on the compactness of $SU(2)$.
In the Minkowskian case the gauge group that will be induced
on the boundary is the left handed part of the Lorentz group,
which is not compact.  However,  it may be 
possible to extend some of the results by finding an appropriate
notion of analytic continuation.

Some comments may be made about the appearance of quantum
groups in the formalism of quantum gravity.  How the loop
algebra in $3+1$ quantum gravity may in fact be deformed so that
the state space has a basis given by quantum spin networks will
be explained elsewhere\cite{sethlee-qnet}.  But perhaps some
possible implications of the existence of this structure may be
mentioned here.  As has been mentioned 
before \cite{spinnet-us}
at finite $k$ there is an infared cutoff in the spectra of
diffeomorphism invariant states of the gravitational field.  It is
then not surprising that this is associated with a finite cosmological,
which is inversely proportional to $k$, as we find here.  (It
is suggestive also  that
Carlip finds a similar relation in $2+1$ dimensions \cite{carlip}.)
It is possible that this infared cutoff may play a useful role
in the statistical mechanics of the spin network states aimed at
understanding quantum gravity at finite temperature.

But, beyond this, the appearance of quantum spin networks opens
up the possibility of qualitatively new phenomena in quantum
gravity.  One is the following.  The states we are speaking of
here are closely connected to the Jones polynomial, especially
if the conjecture about the triviality of the physical kernals is
correct.  We may note that in such states there emerges a
non-trivial duality between topology and geometry.  To see this,
consider a quantum state of the type we are describing here, 
associated to the Jones polynomial.  Let us consider the evaluation
of the state on some spin network of the form of 
$\Gamma \cup \alpha_j$, where $\alpha$ is a component that does
not intersect with the rest, but may be knotted and linked to the rest. 
The $j$ labels the representation associated with the
closed loop $\alpha$.
We assume that this is being evaluated in some
three manifold with topology $\Sigma$.
 For every representation $j$ labeling the loop $\alpha$,
there is a value of the state given by 
$\Psi^J [\Gamma \cup \alpha_j, \Sigma] $.  
Now, as we have described, $j$
has a geometrical interpretation, it tells us how much area there
may be in any surface that intersects only the $\alpha$ loop
once.  Now, it is known that there are coefficients,
$C_j$, such that\cite{witten-tqft} 
\f
\Psi^J [\Gamma , \Sigma^\prime ] =\sum_j c_j
\Psi^J [\Gamma \cup \alpha_j, \Sigma] 
\ff
where $\Sigma^\prime$ is a different three manifold which
is constructed by cuting out from $\Sigma$ a torus around the loop 
$\alpha$ ,
twisting it and then identifying the surfaces of the torus we
removed.

The meaning of this formula is that the value of the state on
the spin network $\Gamma$ in one topology is related to a sum
over its values, on  different spin networks in a different topology,
which differ from each other only by the eigenvalue of a geometric
observable, the areas of surfaces pierced by $\alpha$.
This suggests the existence of a kind of duality between geometry
and topology.

This, of course, depends on the assumption that the physical
states are those which are related to the Jones polynomial.
This is the most important problem not
solved here: to characterize 
the solutions of the quantum dynamics  in the
presence of the boundary conditions.  We have conjectured that
there will be only one unique solution to the hamiltonian
constraint that matches each possible state in a Chern-Simon
theory associated with a set of punctures and representations.
Related to this, as well, is the question
of whether the hamiltonian defined on the boundary in section
Vd. can be defined quantum mechanically acting on the states
we have described here.

We may note, finally, that if this conjecture about the
triviality of the physical kernels is true, the results here may
be considered to constitute a solution of quantum euclidean general
relativity in the case with these boundary conditions, in
terms of the infinite dimensional state space (35) and
the realization on it, of the observable algebra 
${\cal A}^{boundary}_{\cal S}$.  The investigation of this
conjecture is then a priority for further work.

\section*{ACKNOWLEDGEMENTS}

This work has grown out of collaborations with Carlo Rovelli
and was also motivated by conversations with Louis Crane.
In addition, I am grateful to Abhay Ashtekar, John Baez, 
Roumen Borissov, Steve Carlip,  Ted Jacobson, 
Louis Kauffman,
Seth Major, Roger Penrose, Jorge Pullin, Chopin
Soo, Leonard Susskind and
Gerard 't Hooft for conversations and
suggestions in the course of this work.    This work was 
begun during a visit to SISSA in Trieste and completed
during a visit to the Institute for Advanced Study in Princeton, 
both of which were most helpful.  This work
has been supported by the NSF grant  PHY-90-12099.


\begin{thebibliography}{99}
\bibitem{other-categorists}Category-theoretic formulations
of topological quantum field theories have also been
explored in Baez, in this volume and Lawrence,
\end{thebibliography}
\end{document}